\documentclass[pra,aps,amsmath,amssymb,amsfonts,twocolumn,nofootinbib,longbibliography]{revtex4-1}
\usepackage{}
\usepackage{amsmath}    
\usepackage{amssymb}    
\usepackage{amsfonts}   
\usepackage{mathrsfs}   
\usepackage{bbm}        
\usepackage{bm}         
\usepackage{booktabs}   
\usepackage{diagbox}    
\usepackage{multirow}   
\usepackage{graphicx}   
\usepackage{float}      
\usepackage{tipa}       
\usepackage{times}      
\usepackage{verbatim}   
\usepackage[sort&compress]{natbib} 

\allowdisplaybreaks[4]

\usepackage[
  colorlinks,
  breaklinks,
  linkcolor=blue,
  anchorcolor=blue,
  citecolor=blue,
  urlcolor=magenta
]{hyperref}

\begin{document}
\title{Frequency conversion between optical and microwave photons in non-Markovian environments}
\author{Jia Tang$^{1}$ and H. Z. Shen$^{1,2,}$}
\email{Corresponding author: shenhz458@nenu.edu.cn}

\affiliation{$^1$Center for Quantum Sciences and School of Physics,
Northeast Normal University, Changchun 130024, China\\
$^2$Center for Advanced Optoelectronic Functional  Materials
Research, and Key Laboratory for UV Light-Emitting Materials and
Technology of Ministry of Education, Northeast Normal  University,
Changchun 130024, China}

\date{\today}

\begin{abstract}

In this paper, we propose a scheme for frequency conversion between optical photons and microwave photons in non-Markovian environments using both magnetic and mechanical excitations as intermediate media. When the frequencies of optical photons, magnons, phonons, and microwave photons resonance, the conversion efficiency can be made close to reach 98.76$\%$ by adjusting the defined complex cooperativities, while in the case of Markovian, the conversion efficiency is 90.44$\%$. By controlling the environmental spectral widths, the efficiency of frequency conversion exhibits a transition from Markovian regimes to non-Markovian regimes. This transformation simultaneously improves frequency conversion efficiency and conversion bandwidth, which is due to the excitation backflow generated by the interaction between the system and the non-Markovian environments. In the case, when the optical pump power in the non-Markovian regimes are of a large order of magnitude, the conversion bandwidth can be increased, but at the cost of reduced conversion efficiency. Our scheme improves the frequency conversion efficiency and bandwidth between optical photons and microwave photons, breaking the limitations of frequency conversion in Markovian environments and providing a new approach for long-distance quantum communication research of other non-Markovian quantum systems in quantum optics.
\end{abstract}

\maketitle

\section{Introduction}
Photons at optical and microwave frequencies are the most widely deployed signal mediums in the field of modern quantum information processing and communication systems \cite{Nielsen2000,Sergienko2018}. Optical photons can be transmitted by fiber \cite{BonsmaFisher1292036032022,Davtyan1221439022019}, therefore they are suitable for transmitting quantum information which can propagate over long distances. Unlike optical photons, microwave photons can be effectively controlled at the local circuit level. Over the past decades, there have been several excitations exploiting at optical frequencies, including trapped ions \cite{Walker1202036012018}, neutral atoms \cite{Matsukevich3066636662004}, quantum dots \cite{Gao4914264302012,Greve4914214252012}, solid-state defects \cite{Togan4667307342010,Koehl47984872011}, and others operating at microwave frequencies including superconducting qubits \cite{Girvin2014,Kjaergaard113692020} and spins in crystals \cite{Zhong8200320152019}. In addition, many schemes have been investigated to boost the coupling with a wide range of nonlinear frequency mixing mechanisms, including optomechanics, electro-optics, optomagnonics, solid-state spins, trapped atoms/ions, etc. The availability of telecom band single photon detectors \cite{Fisher1270236022021}, quantum memories \cite{Lvovsky37067142009}, and other technologies common in quantum optics experiments \cite{Bagci50781852014} also suggests the need for the development of techniques in the bidirectional transfer of quantum information between optical and microwave photons. These requirements combine to make the task a challenging one.

The frequency conversion of optical and microwave photons needs to involve nonlinear processes to compensate for the large energy difference between optical and microwave photons because the direct microwave and optical photons coupling is extremely weak. Despite a frequency difference of five orders of magnitude, there has been significant progress recently toward the transfer between optical and microwave photons with steadily improved efficiency in a coherent \cite{Peairs140610012020,Lambert319000772020} and bidirectional manner. Nonlinearities play crucial roles in optical and microwave photons frequency conversion because it allows the use of an external pump to compensate for the energy difference between the optical and microwave modes, meanwhile maintaining the phase coherence between input and output signals. The nonlinearity can also emerge due to indirect coupling mediated by another mode, such as mechanical or piezoelectric vibrations or magnetostatic modes. Mechanical resonators have been realized in optomechanical crystals \cite{Eichenfield462782009} interacting with microwave and optical synchronous vibration modes through optomechanical interactions.

For the magnetic excitation, the yttrium iron garnet (YIG) is treated as an intermediate to interact with the optical and microwave modes simultaneously. YIG has attracted great attention and made significant progress over the past decade \cite{LachanceQuirion120701012019,Li1281309022020,Yuan96512022,Rameshti97912022,Zheng1341511012023,Xie250730092023} (see these relevant reviews for more details). In particular, the successful experimental realization of the strong coupling between microwave photons and magnons in a YIG crystal \cite{Tabuchi1130836032014,Zhang1131564012014,Goryachev20540022014}, as theoretically predicted \cite{OOSoykal1040772022010,OOSoykal821044132010}, has kick-started active research in the emerging field of cavity magnonics \cite{Rameshti97912022}. Due to relatively low damping rate,  the spin excitation stays coherent for a time long enough to be able to strongly interact with a microwave photon mode resulting in the hybridized eigenmodes of the coupled systems \cite{Tabuchi1130836032014,Zhang1131564012014}. Strong coupling \cite{Zhang1131564012014,Zhu1010438422020} and even ultrastrong coupling \cite{FornDiaz910250052019,Bourhill931444202016,Kostylev1080624022016} between magnons and microwave cavity photons has been demonstrated in YIG devices. Experimentally, significant progress has been made based on hybrid cavity magnonic systems for many quantum technological applications \cite{Wang1200572022018,Wang942244102016,Bi1272239092020,Crescini1171440012020,Wang1231272022019,Xu1000944152019,Crescini1040644262021,Rao210650012019}. Conversion efficiency is mainly limited by the weak light-magnon coupling \cite{Hisatomi931744272016}, which could be enhanced by placing optical cavity around the YIG crystal. Moreover, YIG being put in a homogeneous external magnetic field shows distinctive resonance modes for the magnetic (spin) excitations perpendicular to the bias field. The magnetic excitations in the YIG sphere can lead to the geometric deformation of the surface. 
Besides, magnons and phonons can also serve as intermediate excitations to achieve optics-microwave multistage frequency transduction \cite{Engelhardt180440592022,Fan1722008662023,Fan1050335072022,Shen1292436012022}. In the light of these advances, a series of theoretical works have been proposed for realizing quantum entanglement \cite{Li1212036012018,Zhang10230212019,Li210850012019,Yu1242136042020,Luo4610732021,Yang30231262021,Nair1170840012020,Li60240052021}, non-Hermitian effects \cite{Koch40131132022,Kotz50330432023}, phonon laser \cite{Ding9157232019,Xu1030535012021}, magnon blockade \cite{Kun1010423312020,Liu1001344212019,Zhao1010638382020,Wu1030524112021}, nonreciprocity \cite{Kong120340012019,Ren1050137112022,EshaqiSani1060326062022}, etc.

The study of open quantum systems is of great significance to understanding and exploring the properties of quantum systems subjected to noises \cite{Breuer2002,Weiss2008,Gardiner2000,Li810621242010,Franco2713450532013,Vega890150012017}. In fact, all quantum systems in practical applications are open due to the interaction of systems with the external environments. The Markovian approximation is valid when the interacting strengths between the quantum systems and the environments are weak and the characteristic times of the studied systems are much larger than those of the baths. In general, we should consider non-Markovian excitation backflow effects of the multiple environments to systems \cite{Breuer700453232004,Ferraro800421122009,Shen1070537052023,Chan890421172014,Man920123152015}, which include quantum state engineering, quantum feedback control \cite{Xue860523042012}, and quantum channel capacity \cite{Bylicka457202014}. Non-Markovian effects occur in many quantum systems, including coupled cavities \cite{Link30203482022}, photonic crystals \cite{Burgess1050622072022Hoeppe1080436032012}, color noises \cite{CostaFilho950521262017}, the cavity and atom coupled to waveguides \cite{Tan830321022011,Xin1050537062022}, which have been experimentally implemented \cite{Liu79312011,Xiong1000321012019,Cialdi1000521042019,Groblacher676062015,Khurana990221072019,Madsen1062336012011,Guo1262304012021,Li1291405012022Xu820423282010Tang97100022012,Uriri1010521072020,Anderson4732021993,Liu1020622082020,Fanchini1122104022014,Haseli900521182014,Goswami1040224322021,Debiossac1282006012022}.
The excitation backflow between the systems and their environments can characterize the non-Markovian effects of the environments back-acting on the systems dynamics \cite{Breuer880210022016,Breuer1032104012009,Wibmann860621082012,Wibmann920421082015} with various measures of non-Markovianity \cite{Lorenzo88020102R2013,Rivas1050504032010,Luo860441012012,Wolf1011504022008,Lu820421032010,Chruscinski1121204042014,Hou830621152011,Hou860121012012}.

Inspired by Ref.~\cite{Engelhardt180440592022}, which provided valuable insights into frequency conversion between optical and microwave photons processes in Markovian environments, we are motivated to explore the uncharted territory of frequency conversion between optical and microwave photons in non-Markovian environments. This research is based on the work of Ref.~\cite{Engelhardt180440592022}. By doing so, we aim to systematically contrast and analyze the differences in frequency conversion between optical and microwave photons characteristics between Markovian and non-Markovian environments. This approach not only allows for a comprehensive understanding of the influences of non-Markovian effects on frequency conversion but also provides a novel perspective for the further development of this field.

The above considerations motivate us to investigate these problems: (i) How can we construct the model in non-Markovian environments to the two-stage optics-to-microwave conversion systems? (ii) Can conversion efficiency and bandwidths be enhanced by generalizing frequency conversion from Markovian systems to non-Markovian ones? (iii) How can non-Markovian effects affect the conversion efficiency, conversion bandwidths, and frequency spectrum of the resonances for the interacting systems? To address these problems, we propose a scheme to realize a two-stage frequency conversion in which both mechanical excitation and magnetic excitation as the intermediate medium between optical and microwave photons in non-Markovian environments. 

This paper is organized as follows. In Sec.~II, we first introduce the model Hamiltonian. In Sec.~III, the exact non-Markovian input-output relations are derived, which can return to the Markovian input-output ones in the wideband limit. We set up a two-stage frequency conversion device based on magnons and phonons as intermediate media in non-Markovian environments, and calculate the conversion efficiency. In Sec.~IV, we discuss the conversion efficiency and the bandwidths with non-Markovian effects. The main conclusions are summarized in Sec.~V.
\section{MODEL AND HAMILTONIAN}
\begin{figure}[h]
\centering
\includegraphics[angle=0,width=0.45\textwidth]{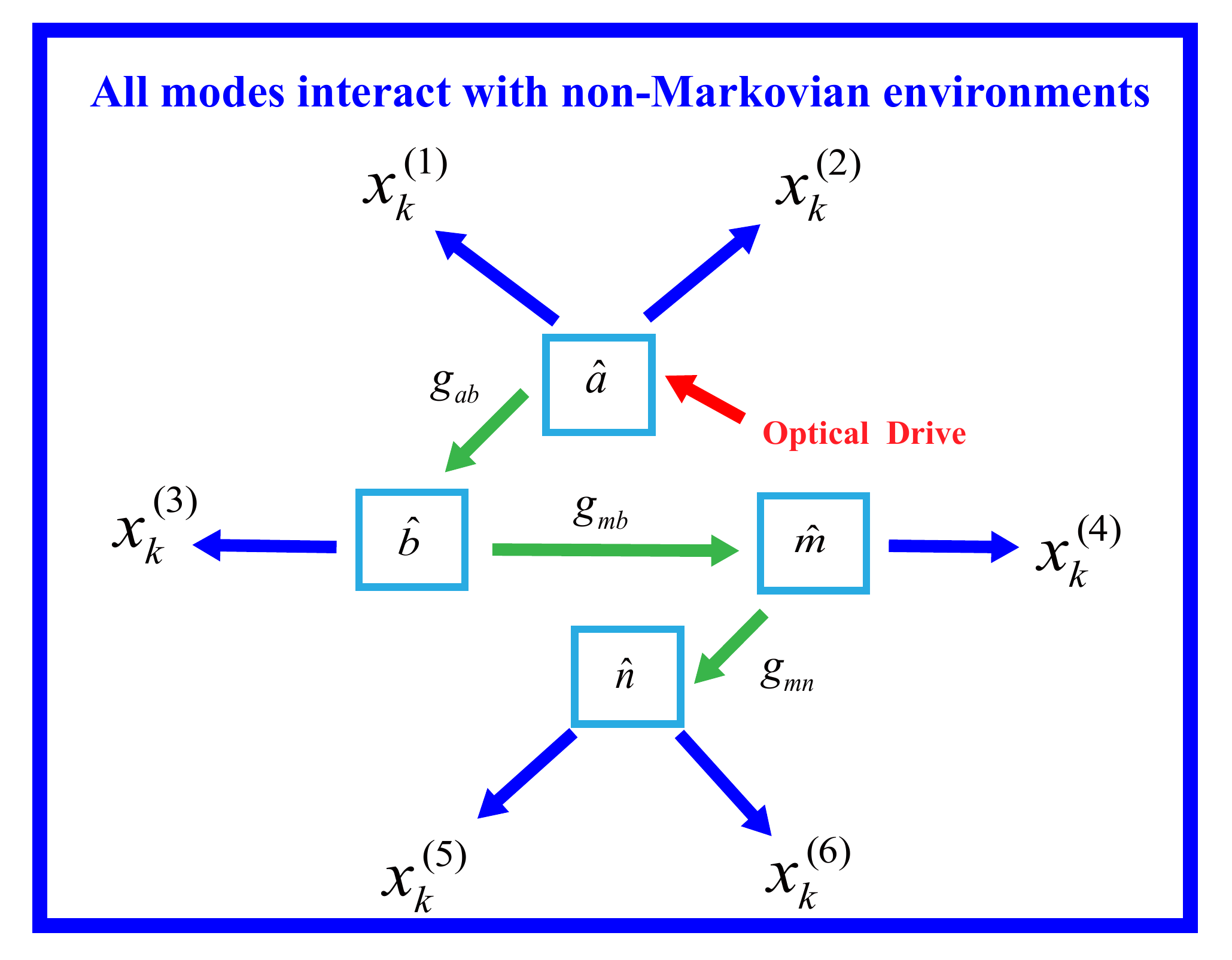}
\caption{(Color online)   A sketch of a two-stage frequency conversion  device in non-Markovian environments. The two intermediate media between optical photons ($\hat a$) and microwave photons ($\hat n$) are phonons ($\hat b$) and magnons ($\hat m$), which have frequencies both within the microwave range. The optical photon mode interacts with two non-Markovian environments. Two non-Markovian environments are added to the microwave photon mode. Phonon mode and magnon mode each interact with their respective non-Markovian environment. 
The red arrow indicates the optical drive. The thick lines denote the dominant interaction scheme.}
\label{xinmoxing}
\end{figure}

We consider a hybrid quantum system consisting of four bosonic modes with six non-Markovian environments sketched in Fig.~\ref{xinmoxing}, where optical photons and microwave photons each interact with two non-Markovian environments, while phonons and magnetons each interact with one non-Markovian environment. 

The total Hamiltonian is given by \cite{Engelhardt180440592022}
\begin{small}
\begin{equation}
\begin{aligned}
{\hat H_1} = {\hat H_S} + {\hat H_D} + {\hat H_E},
\label{eq1}
\end{aligned}
\end{equation}
\end{small}with the systems parts
\begin{small}
\begin{equation}
\begin{aligned}
{{\hat H}_S}/\hbar = &{\omega_a}{{\hat a}^\dag }\hat a + {\omega_b}{{\hat b}^\dag }\hat b + {\omega_m}{{\hat m}^\dag }\hat m + {\omega_n}{{\hat n}^\dag }\hat n\\
&+ {\textsl{g}_{ab}}{{\hat a}^\dag }\hat a({{\hat b}^\dag } + \hat b) + {\textsl{g}_{mb}}({{\hat b}^\dag } + \hat b)({{\hat m}^\dag } + \hat m)\\
&+ {\textsl{g}_{mn}}({{\hat m}^\dag } + \hat m)({{\hat n}^\dag } + \hat n),
\label{eq2}
\end{aligned}
\end{equation}
\end{small}the optical drive term
\begin{small}
\begin{equation}
\begin{aligned}
{\hat H_D}/\hbar = & i\mathbbm{a} (\hat a{e^{i{\omega_D}t}} - {\hat a^\dag }{e^{ - i{\omega_D}t}}),
\label{eq4}
\end{aligned}
\end{equation}
\end{small}and the non-Markovian environments parts
\begin{small}
\begin{equation}
\begin{aligned}
{{\hat H}_E}/\hbar = \sum\limits_{\nu,k} {c_k'^{(\nu)} } \hat d_k^{{(\nu)} \dag }\hat d_k^{(\nu)} + \sum\limits_{\nu,k} {(x_k^{(\nu)} } { \hat y_\nu }\hat d_k^{{(\nu)} \dag } + x_k^{{(\nu)} *}\hat y_\nu ^\dag \hat d_k^{(\nu)}),
\label{eq3}
\end{aligned}
\end{equation}
\end{small}where ${\hat a}^\dag$ ($\hat a$) denotes the creation (annihilation) operators of the optical photon mode with frequency $\omega_a$. ${\hat b}^\dag$ ($\hat b$) describes the creation (annihilation) operator of the phonon mode with frequency $\omega_b$. ${\hat m}^\dag$ ($\hat m$) represents the creation (annihilation) operator of the magnon mode with frequency $\omega_m$. ${\hat n}^\dag$ ($\hat n$) is the creation (annihilation) operator of the microwave photon mode with frequency $\omega_n$. ${\textsl{g}_{ab}}$, ${\textsl{g}_{mb}}$, and ${\textsl{g}_{mn}}$ correspond to the optomechanical, magnomechanical, and magnon-microwave interacting strengths, respectively. The bosonic modes of the systems interact with the $k$th mode (eigenfrequencies ${c_k'^{(\nu)} }$) of non-Markovian environments, which are modeled as collections of infinite modes via the creation (annihilation) operators $\hat d_k^{{(\nu)}\dag } $ ($\hat d_k^{(\nu)}$). The parameter ${x_k^{(\nu)} }$ is interacting strength between six environments and four bosonic modes (including the optical photon mode, phonon mode, magnon mode, and microwave photon mode). In Eq.~(\ref{eq3}), we have defined $\nu=1,2,3,4,5,6$, ${{\hat y}_1} = {\hat y}_2 = \hat a$, ${{\hat y}_3} = \hat b$, ${\hat y}_4 = \hat m $, and ${\hat y}_5 = {{\hat y}_6} = \hat n.$
\begin{table}[h]
\centering
\caption{Table of studied parameters for different mode frequencies, linewidths or dissipations and interacting strengths \cite{Engelhardt180440592022}.}
\label{Table1}
   \begin{tabular}{lcc}
       \hline
       \hline
        {Quantity}&{Symbol}&{Value}\\
       \hline
        {Optical photon frequency}&{${\omega _a}/2\pi$}&{200 THz}\\
        {Optical detuning(red)}&{${\Omega _a}/2\pi$}&{-10 GHz}\\
        {Phonon frequency}&{${\omega _b}/2\pi$}&{10 GHz}\\
        {Magnon frequency }&{${\omega _m}/2\pi$}&{10 GHz}\\
        {External optical dissipation}&{${\Gamma _1}/2\pi$}&{1.5 GHz}\\
        {Internal optical linewidth}&{${\Gamma _2}/2\pi$}&{0.1 GHz}\\
        {Phonon linewidth}&{${\Gamma _3}/2\pi$}&{1.0 kHz}\\
        {Magnon linewidth}&{${\Gamma _4}/2\pi$}&{1.0 MHz}\\
        {External microwave cavity interacting strength}&{${\Gamma _5}/2\pi$}&{100 MHz}\\
        {Internal microwave cavity loss strength}&{${\Gamma _6}/2\pi$}&{3.0 MHz}\\
        {Magnomechanical interacting strength}&{${\textsl{g}_{mb}}/2\pi $}&{10.0 MHz}\\
        {Magnon-microwave photon interacting strength}&{${\textsl{g}_{mn}}/2\pi $}&{200 MHz}\\
        {Optomechanical interacting strength}&{${\textsl{g}_{ab}}/2\pi $}&{0.2MHz}\\
       \hline
       \hline
     \end{tabular}
\end{table}

After a unitary transformation via $\hat U\hat H_1{{\hat U}^\dag } - i\hbar \hat U \partial {\hat U^\dag }/\partial t $ with $\hat U = \exp [i{\omega_D}t({{\hat a}^\dag }\hat a+\sum\nolimits_k \hat d_k^{{(1)} \dag }\hat d_k^{(1)}+\sum\nolimits_k \hat d_k^{{(2)} \dag }\hat d_k^{(2)})]$ to the rotating frame, we linearize the Hamiltonian (\ref{eq1}) by taking into account fluctuations on top of the drive-induced coherent steady state. Only the optical field shows a coherent part for small interacting strengths, as shown by $\hat a = \mathbbm{a}  + \delta \hat a$ with $\mathbbm{a}  = \sqrt {N_a} $, where ${N_a}$ is the mean photon number in the optical cavity. 
${\Gamma _1}$ denotes the dissipation strength to the external port. ${\Gamma _2}$ represents the internal optical mode linewidth. ${\Omega _a}={\omega _D} - {\omega _a}$ describes the detuning from the drive frequency. ${c_k^{(1)}}={c_k'^{(1)}}-{\omega _D}$ and ${c_k^{(2)}}={c_k'^{(2)}}-{\omega _D}$ are the detunings from the two non-Markovian environments and optical mode, where we have defined ${c_k^{(3)}}={c_k'^{(3)}}$, ${c_k^{(4)}}={c_k'^{(4)}}$, ${c_k^{(5)}}={c_k'^{(5)}}$, and ${c_k^{(6)}}={c_k'^{(6)}}$. The confinement of photons yields a interacting enhancement of the parametric processes that scales with the square root of the mean photon number, where the cavity enhanced interacting strength is given by ${G_{ab}} = \sqrt {N_a} {\textsl{g}_{ab}}$ \cite{Aspelmeyer8613912014, Engelhardt180440592022}.

With the rotating-wave approximation, 
we obtain the effective Hamiltonian
\begin{small}
\begin{equation}
\begin{aligned}
{\hat H_2} /\hbar=& - {\Omega _a}{{\hat a}^\dag }\hat a + {\omega_b}{{\hat b}^\dag }\hat b + {\omega_m}{{\hat m}^\dag }\hat m + {\omega_n}{{\hat n}^\dag }\hat n\\
             &+ {G_{ab}}({{\hat a}^\dag }\hat b+ \hat a {{\hat b}^\dag }) + {\textsl{g}_{mb}}({{\hat m}^\dag }\hat b+ \hat m {{\hat b}^\dag })\\
              &+ {\textsl{g}_{mn}}({{\hat m}^\dag }\hat n+ \hat m {{\hat n}^\dag })\\
              &+\sum\limits_{\nu,k} {c_k^{(\nu)} } \hat d_k^{{(\nu)} \dag }\hat d_k^{(\nu)} + \sum\limits_{\nu,k} {(x_k^{(\nu)} } { \hat y_\nu }\hat d_k^{{(\nu)} \dag } + x_k^{{(\nu)} *}\hat y_\nu ^\dag \hat d_k^{(\nu)}),
\label{eq9}
\end{aligned}
\end{equation}
\end{small}
whose derivation can be found in Appendix \ref{A}.

\section{CONVERSION EFFICIENCY WITH NON-MARKOVIAN EFFECTS}
With the operators expectation values defined by $a(t) \equiv \langle \hat a(t) \rangle $, $b(t) \equiv \langle \hat b(t) \rangle $, $m(t) \equiv \langle \hat m(t) \rangle $, $n(t) \equiv \langle \hat n(t) \rangle $, $d_k^{(\nu)} (t) \equiv \langle {\hat d_k^{(\nu)}}(t)\rangle $, the Heisenberg equations $\dot {\hat A}(t) =  - i[\hat A(t),{{\hat H}}(t)]$ of the systems give
\begin{small}
\begin{equation}
\begin{aligned}
{\dot a}(t) = &  i{\Omega _a} a(t) - i{G_{ab}} b(t) - i\sum\limits_k x_k^{{(1)}*}d_k^{(1)}(t)\\
&- i\sum\limits_k x_k^{{(2)}*}d_k^{(2)}(t),\\
{\dot b}(t) =  & - i{\omega_b} b(t) - i{G_{ab}} a(t)- i{\textsl{g}_{mb}} m(t) - i\sum\limits_k x_k^{{(3)}*}d_k^{(3)}(t),\\
{\dot m}(t) =& - i{\omega_m} m(t) - i{\textsl{g}_{mb}} b(t)- i{\textsl{g}_{mn}} n(t)- i\sum\limits_k x_k^{{(4)}*}d_k^{(4)}(t),\\
{\dot n}(t) = &  - i{\omega_n} n(t) - i{\textsl{g}_{mn}} m(t)-i\sum\limits_k x_k^{{(5)}*}d_k^{(5)}(t)\\
&  - i\sum\limits_k x_k^{{(6)}*}d_k^{(6)}(t),\\
\dot d_k^{(\nu)} (t) = & - i{c_k^{(\nu)}} {d_k^{(\nu)}}(t) - i {x_k^{(\nu)} }{{ y}_\nu }(t).
\label{eq10}
\end{aligned}
\end{equation}
\end{small}Through simple calculations by solving Eq.~(\ref{eq10}), we get solutions of the environments parts for $t \ge 0$
\begin{small}
\begin{equation}
\begin{aligned}
{d_k^{(\nu)}}(t) ={d_k^{(\nu)}}(0){e^{ - i{c_k^{(\nu)}}t}} - i{x_k^{(\nu)}}\int_0^t {d\tau {{ y}_\nu }(\tau )} {e^{ - i{c_k^{(\nu)}}(t - \tau )}},
\label{eq12}
\end{aligned}
\end{equation}
\end{small}where we have used the definition $\hat d_k^{(\nu)} (0) \equiv \hat d_k^{(\nu)}$ in the process of deriving Eq.~(\ref{eq12}). The first terms on the right-hand side of Eq.~(\ref{eq12}) represent the free-propagating parts of the environments fields, while the second terms describe the influences of non-Markovian environments on the systems dynamics. Substituting Eq.~(\ref{eq12}) into Eq.~(\ref{eq10}), we can obtain the integro-differential equations
\begin{small}
\begin{equation}
\begin{aligned}
{\dot a}(t) = &i{\Omega _a} a(t) - i{G_{ab}} b(t) - K_1 (t) - K_2 (t) \\
&- \int_0^t {d\tau  a(\tau )} {f_1}(t - \tau ) - \int_0^t {d\tau  a(\tau )} {f_2}(t - \tau ),\\
{\dot b}(t) = & - i{\omega_b} b(t) - i{G_{ab}} a(t) - i{\textsl{g}_{mb}} m(t) -  K_3 (t) \\
&- \int_0^t {d\tau b(\tau )} {f_3}(t - \tau ),\\
{\dot m}(t) = & - i{\omega_m} m(t) - i{\textsl{g}_{mb}} b(t) - i{\textsl{g}_{mn}} n(t) -  K_4 (t) \\
&- \int_0^t {d\tau  m(\tau )} {f_4}(t - \tau ),\\
{\dot n}(t) = & - i{\omega_n} n(t) - i{\textsl{g}_{mn}} m(t) -  K_5 (t) -  K_6 (t)\\
&- \int_0^t {d\tau  n(\tau )} {f_5}(t - \tau ) - \int_0^t {d\tau  n(\tau )} {f_6}(t - \tau ),
\label{eq13}
\end{aligned}
\end{equation}
\end{small}where the externally driven environment functions $K_\nu (t) = i\sum\nolimits_k {x_k^{{(\nu)} *}{{d}_k^{(\nu)}}(0){e^{ - i{c _k^{(\nu)}}t}}}  = \int_{ - \infty }^\infty  {d\tau \kappa _\nu^*} (t - \tau ) $ $ {y}_{in}^{(\nu )} (\tau )$. The systems interact with the incoming and outgoing modes of environments at both ends, where we have defined the expectation values of the input field operators as ${y}_{in}^{(\nu )}(t) = \frac{-1}{{\sqrt {2\pi } }}\sum\nolimits_k {{e^{ - i{{c_k^{(\nu)} }}t}}{d_k^{(\nu)} }(0)} $, where ${y}_{in}^{(1 )}(t)={ a}_{in}^{(1)}(t)$, ${y}_{in}^{(2)}(t)={ a}_{in}^{(2)}(t)$, ${y}_{in}^{(3 )}(t)={ b}_{in}^{(3)}(t)$, ${y }_{in}^{(4 )}(t)={ m}_{in}^{(4)}(t)$, ${y }_{in}^{(5 )}(t)={n}_{in}^{(5)}(t)$, ${y}_{in}^{(6 )}(t)={n}_{in}^{(6)}(t)$. The impulse response functions read ${\kappa _\nu}(t) = \frac{{ i}}{{\sqrt {2\pi } }}\sum\nolimits_k {{e^{i{c _k^{(\nu)}}t}}{x_k^{(\nu)}}}$, or in the continuum
\begin{small}
\begin{equation}
\begin{aligned}
{\kappa _\nu}(t  - \tau) = \frac{{ i}}{{\sqrt {2\pi } }}\int  {{e^{  ic^{(\nu)}(\omega) (t - \tau )}}x^{(\nu)}(\omega )} d\omega,
\label{impulse_response_function}
\end{aligned}
\end{equation}
\end{small}where we have made the replacements by $c_k^{(\nu)}\to c^{(\nu)}(\omega )$ and $x_k^{(\nu)}\to x^{(\nu)}(\omega )$. The memory functions are given by
\begin{small}
\begin{equation}
\begin{aligned}
{f_\nu}(t) = \sum\limits_k {{{\left| {{x_k^{(\nu)}}} \right|}^2}{e^{ - i{c _k^{(\nu)}}t}}}  = \int  {{J_\nu}(\omega ){e^{ - i\omega t}}d\omega },
\label{correlation_function}
\end{aligned}
\end{equation}
\end{small}where ${J_\nu}(\omega ) = {\sum\nolimits_k {{{\left| {{x_k^{(\nu)}}} \right|}^2}} } \delta (\omega  - {c _k^{(\nu)}})$ represents the spectral densities of the environments, respectively.  ${f_\nu}(t)$ denotes the memory functions of the environments, which describe the non-Markovian fluctuation-dissipation relationship of the environments.
In a similar manner, we derive solutions of the environments parts for ${t_1} \ge t$ from Eq.~(\ref{eq10}), i.e.,
\begin{small}
\begin{equation}
\begin{aligned}
{{ d}_k^{(\nu)}}(t) = {d_k^{(\nu)}}({t_1}){e^{ - i{c_k^{(\nu)}}(t - {t_1})}} + i{x_k^{(\nu)}}\int_t^{t_1} {d\tau {y_\nu }(\tau )} {e^{ - i{c_k^{(\nu)}}(t - \tau )}},
\label{eq14}
\end{aligned}
\end{equation}
\end{small}and the integro-differential equations
\begin{small}
\begin{equation}
\begin{aligned}
{\dot a}(t)= &i{\Omega _a} a(t) - i{G_{ab}} b(t) - { K'_1}(t) - {K'_2}(t)\\
&+ \int_t^{{t_1}} {d\tau  a(\tau )} {f_1}(t - \tau ) + \int_t^{{t_1}} {d\tau  a(\tau )} {f_2}(t - \tau ),\\
{\dot b}(t) =  &- i{\omega_b}b(t) - i{G_{ab}} a(t) - i{\textsl{g}_{mb}} m(t) - { K'_3}(t)\\
&+ \int_t^{{t_1}} {d\tau  b(\tau )} {f_3}(t - \tau ),\\
{\dot m}(t) =  &- i{\omega_m} m(t) - i{\textsl{g}_{mb}} b(t) - i{\textsl{g}_{mn}} n(t) - { K'_4}(t) \\
&+ \int_t^{{t_1}} {d\tau  m(\tau )} {f_4}(t - \tau ),\\
{\dot n}(t) =  &- i{\omega_n} n(t) - i{\textsl{g}_{mn}} m(t) - { K'_5}(t) - { K'_6}(t)\\
&+ \int_t^{{t_1}} {d\tau  n(\tau )} {f_5}(t - \tau ) + \int_t^{{t_1}} {d\tau  n(\tau )} {f_6}(t - \tau ),
\label{eq15}
\end{aligned}
\end{equation}
\end{small}where the externally driven environment functions ${ K'_\nu}(t) = i\sum\nolimits_k x_k^{{(\nu)} *} {d_k^{(\nu)}}({t_1}){e^{ - i{c _k^{(\nu)}}(t - {t_1})}} = \int_{ - \infty }^\infty  {d\tau \kappa _\nu^*} (t - \tau )$\\${y }_{out}^{(\nu )}(\tau )$, with the defined expectation values of the output field operators and the external driven environments operators as ${y}_{out}^{(\nu )}(t) = \frac{1}{{\sqrt {2\pi } }}\sum\nolimits_k {{e^{ - i{c _k^{(\nu)}}(t - {t_1})}}{{ d}_k^{(\nu)}}({t_1})}$, where ${y }_{out}^{(1 )}(t)={ a}_{out}^{(1)}(t)$, ${y}_{out}^{(2)}(t)={ a}_{out}^{(2)}(t)$, ${y}_{out}^{(3 )}(t)={ b}_{out}^{(3)}(t)$, ${y}_{out}^{(4 )}(t)={ m}_{out}^{(4)}(t)$, ${y}_{out}^{(5 )}(t)={n}_{out}^{(5)}(t)$, ${y}_{out}^{(6 )}(t)={n}_{out}^{(6)}(t)$. $\kappa_\nu(t)$ and $f_\nu(t)$ are determined by Eqs.~(\ref{impulse_response_function}) and ~(\ref{correlation_function}).

By comparing Eq.~(\ref{eq13}) with Eq.~(\ref{eq15}), the input and output fields \cite{collett3013861984,gardiner3137611985} are connected by the non-Markovian input-output relations \cite{shen880338352013,diosi850341012012} and systems parts (setting ${t_1} \to t$)
\begin{small}
\begin{equation}
\begin{aligned}
{y }_{out}^{(\nu )}(t) + {y }_{in}^{(\nu )}(t) = -\int_{{0}}^t {d\tau {\kappa _\nu }(\tau - t)} {{ y}_\nu }(\tau ),
\label{nonMarkovian_input_output}
\end{aligned}
\end{equation}
\end{small}where the concrete form of ${\kappa _\nu }(t - \tau)$ is given by Eq.~(\ref{impulse_response_function}). The spectral response functions are assumed as \cite{shen880338352013,Shen1070537052023,shen1050237072022,Zhang1090337012024,Yang1090537122024} 
\begin{small}
\begin{equation}
\begin{aligned}
x^{(\nu)}(\omega ) = \sqrt {\frac{{{\Gamma _\nu}}}{{2\pi }}} \frac{{{\lambda _\nu}}}{{{\lambda _\nu} - i\omega }},
\label{spectral_response_function}
\end{aligned}
\end{equation}
\end{small}where $\lambda_\nu$ is the environmental spectrum width, while $\Gamma_\nu$ defines the internal linewidths of the  environmental modes.
Thus the spectral density of the $\nu$th environment reads \cite{Breuer2002,xiong860321072012,shen950121562017,shen1050237072022,shen880338352013,zhang870321172013, Jack630438032001,Shen1070537052023,Zhang1090337012024,Yang1090537122024}
\begin{small}
\begin{equation}
\begin{aligned}
{J_\nu }(\omega ) = \frac{{{\Gamma _\nu }}}{{2\pi }}\frac{{\lambda _\nu ^2}}{{\lambda _\nu ^2 + {\omega ^2}}},
\label{spectral_density}
\end{aligned}
\end{equation}
\end{small}which corresponds to the Lorentzian spectral density. Specifically, the parameter $\lambda_\nu$ is inversely proportional to the environmental correlation time. Moreover, the Lorentzian spectral density in Eq.~(\ref{spectral_density}) can also be realized an all-optical setup in \cite{Xiong1000321012019,Cialdi1000521042019,Haseli900521182014,Li1291405012022Xu820423282010Tang97100022012,Fanchini1122104022014} and pseudomode theory \cite{Jack630438032001,Barnett1997, Garraway5522901997,Garraway5546361997,Man900621042014,Man2357632015,Mazzola800121042009,Pleasance960621052017}. With Eqs.~(\ref{spectral_response_function}) and~(\ref{spectral_density}), we have ${\kappa _{\nu} }(\tau  - t) = i\sqrt {{\Gamma _{\nu}}} {\lambda _{\nu}}{e^{ - {\lambda _{\nu}}( {t - \tau })}} \theta (t- \tau)$ and ${f _{\nu}}(t  - \tau) = \frac{1}{2} {\Gamma _{\nu}} {\lambda _{\nu}}{e^{ - {\lambda _{\nu}}\left| {t - \tau } \right|}} $, which represents a Gaussian Ornstein-Uhlenbeck process \cite{uhlenbeck368231930,gillespie5420841996,jing1052404032010}, where $\theta (t - t')$ is the unit step function, and then $\theta (t-t') = 1$ for $t \ge t'$ otherwise $\theta (t-t')=0$.
When $\lambda_\nu$ tends to infinity, the environments become memoryless. That is to say, in the wideband limit (i.e., ${\lambda _\nu } \to \infty $), the spectral density approximately takes ${J_\nu }(\omega ) \to {{\Gamma _\nu }}/{{2\pi }}$, or, equivalently, $x^{(\nu)}(\omega ) \to \sqrt {{{\Gamma _\nu}}/{{2\pi }}}$. This describes the case in the Markovian limit. According to Eqs.~(\ref{impulse_response_function}) and ~(\ref{correlation_function}), we have $f_\nu(t) = \Gamma_\nu \delta(t)$ and ${\kappa _\nu}(t) =   i\sqrt {{\Gamma _\nu}} \delta (t)$. Substituting these results into Eq.~(\ref{nonMarkovian_input_output}), we can obtain the Markovian input-output relations
\begin{small}
\begin{equation}
\begin{aligned}
{y }_{out}^{(\nu )}(t) + {y}_{in}^{(\nu )}(t) = -i\sqrt {{\Gamma _\nu}} {y_\nu }(t).
\label{Markovianinputoutput}
\end{aligned}
\end{equation}
\end{small}We show that the Markovian input-output relations given by Eq.~(\ref{Markovianinputoutput}) are equivalent to those defined in Refs.\cite{Gardiner2000,Walls1994,Scully1997} and can return to the results ${y }_{out}^{(\nu )}(t) - {y}_{in}^{(\nu )}(t) =  \sqrt {{\Gamma _\nu}} {y_\nu }(t)$ of Refs.\cite{Gardiner2000,Walls1994,Scully1997} by the replacements ${x_k^{(\nu)}} \to i{x_k^{(\nu)}}$ (${x^{(\nu)}(\omega)} \to i{x^{(\nu)}(\omega)}$) in Eqs.~(\ref{eq9}),~(\ref{impulse_response_function}), and (\ref{Markovianinputoutput}).

Making a modified Laplace transformation \cite{Uchiyama800211282009,Saeki810311312010,Shen920521222015,shen950121562017} $\zeta(\omega ) = \int_0^\infty  {{e^{i\omega t }}} \zeta(t)dt$ to Eq.~(\ref{eq13}), where ${{e^{i\omega t }}}\to {e^{i\omega t - \epsilon t}}$ with $ \epsilon  \to {0^ + }$ makes $\zeta(\omega)$ converge to a finite value, then the systems equations satisfy
\begin{small}
\begin{equation}
\begin{aligned}
 - i\omega  a(\omega ) = &  i \Omega_a  a(\omega)-i{G_{ab}} b(\omega)\\
 &- {{\tilde \kappa }_1}(\omega )[{ a}_{in}^{(1)}(\omega )-{ a}_{in}^{(1)}(i{\lambda _1})]-  a(\omega ){f_1}(\omega ) \\
&- {{\tilde \kappa }_2}(\omega )[{ a}_{in}^{(2)}(\omega )-{ a}_{in}^{(2)}(i{\lambda _2})]-  a(\omega ){f_2}(\omega ),\\
- i\omega  b(\omega ) = & - i \omega_b  b(\omega)-i{G_{ab}} a(\omega)-i{\textsl{g}_{mb}} m(\omega)\\
&- {{\tilde \kappa }_3}(\omega )[{ b}_{in}^{(3)}(\omega )-{ b}_{in}^{(3)}(i{\lambda _3})]-  b(\omega ){f_3}(\omega ),\\
- i\omega  m(\omega ) = & - i \omega_m  m(\omega)-i{\textsl{g}_{mb}} b(\omega)-i{\textsl{g}_{mn}} n(\omega)\\
&- {{\tilde \kappa }_4}(\omega )[{ m}_{in}^{(4)}(\omega )-{ m}_{in}^{(4)}(i{\lambda _4})]-  m(\omega ){f_4}(\omega ),\\
- i\omega  n(\omega ) =&  - i \omega_n  n(\omega)-i{\textsl{g}_{mn}} m(\omega)\\
 &- {{\tilde \kappa }_5}(\omega )[{ n}_{in}^{(5)}(\omega )-{ n}_{in}^{(5)}(i{\lambda _5})]-  n(\omega ){f_5}(\omega ) \\
&- {{\tilde \kappa }_6}(\omega )[{ n}_{in}^{(6)}(\omega )-{ n}_{in}^{(6)}(i{\lambda _6})]-  n(\omega ){f_6}(\omega ),\\
\label{relation1}
\end{aligned}
\end{equation}
\end{small}with ${{\tilde \kappa }_\nu}(\omega ) = \int_{-\infty}^ 0 {\kappa _\nu^*} (t'){e^{i\omega t' }}dt'$, ${y}_{in}^{(\nu )}(\omega)=\int_0^\infty  {{{y}_{in}^{(\nu )}}} (t'){e^{i\omega t' }}dt'$, and ${f_\nu}(\omega ) = \int_0^\infty  {{f_\nu}} (t'){e^{i\omega t' }}dt'$.

In order to distinguish the effects of the inhomogeneous terms ${ a}_{in}^{(1)}(i{\lambda _1})$, ${ a}_{in}^{(2)}(i{\lambda _2})$, ${ b}_{in}^{(3)}(i{\lambda _3})$, ${ m}_{in}^{(4)}(i{\lambda _4})$, ${ n}_{in}^{(5)}(i{\lambda _5})$, and ${ n}_{in}^{(6)}(i{\lambda _6})$ appearing in Eq.~(\ref{relation1}), taking ${ a}_{in}^{(1)}(t)$ as an example, we define $\phi ({\lambda _1},\omega ) = { a}_{in}^{(1)}(i{\lambda _1})/{ a}_{in}^{(1)}(\omega )$ and  assume that the input field has two concrete forms as follows: damped oscillation ${ a}_{in}^{(1)}(t) = {p}{{\mathop{\rm e}\nolimits} ^{ - \varepsilon t}}\sin ({q}{t^2})$ for $\varepsilon  > 0$ as well as ${q} > 0$ and Gaussian profile $ { a}_{in}^{(1)}(t) = {p}{{\mathop{\rm e}\nolimits} ^{ - \varepsilon t^2}}\cos ({q}{t})$ for $\varepsilon  > 0$ as well as ${q} > 0$, which respectively correspond to $\phi ({\lambda _1},\omega ) = \{ \cos [\frac{{{{\left( {\lambda  + \varepsilon } \right)}^2}}}{{4q}}][1 - 2fc(\frac{{\lambda  + \varepsilon }}{{\sqrt {2\pi q} }})] + [1 - 2fs(\frac{{\lambda  + \varepsilon }}{{\sqrt {2\pi q} }})]\sin [\frac{{{{\left( {\lambda  + \varepsilon } \right)}^2}}}{{4q}}]\} /\{ \cos [\frac{{{{\left( {\varepsilon  - {\rm{i}}\omega } \right)}^2}}}{{4q}}][1 - 2fc(\frac{{\varepsilon  - {\rm{i}}\omega }}{{\sqrt {2\pi q} }})] + [1 - 2fs(\frac{{\varepsilon  - {\rm{i}}\omega }}{{\sqrt {2\pi q} }})]\sin [\frac{{{{\left( {\varepsilon  - {\rm{i}}\omega } \right)}^2}}}{{4q}}]\} $ and $\phi ({\lambda _1},\omega ) = {e^{\frac{{{\lambda ^2} + {\omega ^2} + 2q(\omega  - i\lambda )}}{{4\varepsilon }}}}\{ i + {e^{\frac{{{\rm{i}}q\lambda }}{\varepsilon}}}[i + erfi(\frac{{q - {\rm{i}}\lambda }}{{2\sqrt \varepsilon  }})] - erfi(\frac{{q + {\rm{i}}\lambda }}{{2\sqrt \varepsilon  }})\} /\{ i + {e^{\frac{{q\omega }}{\varepsilon }}}[i + erfi(\frac{{q - \omega }}{{2\sqrt \varepsilon  }})] - erfi(\frac{{q + \omega }}{{2\sqrt \varepsilon  }})\} $, where $fc(\ell ) = \int_0^\ell  {\cos (\pi {t^2}/2)} dt,fs(\ell ) = \int_0^\ell  {\sin (\pi {t^2}/2)} dt,erfi(\ell ) = erf(i\ell )/i $ with $ erf(\ell ) = \frac{2}{{\sqrt \pi  }}\int_0^\ell  {{e^{ - {t^2}}}} dt $. We find that $\phi ({\lambda _1 },\omega )$ is induced by non-Markovian effects and have no Markovian counterparts, which are inhomogeneous terms depending on the specific forms of the input field ${ a}_{in}^{(1)}(t)$. In Markovian approximation, $\phi ({\lambda _1 },\omega )$ tends to zero for ${\lambda _1 } \to \infty $.

We show the inhomogeneous term cannot reveal the characteristics of the systems under probe. For the form of damped oscillation, we can evaluate $\left| {\phi ({\lambda _1},\omega )} \right| \sim {10^{ - 5}}$ for $\lambda  = {\omega _n}$ (falling in non-Markovian regimes), and $ \left| {\phi ({\lambda _1},\omega )} \right| \sim {10^{ - 8}}$ for $\lambda  = 10{\omega _n}$ (weak non-Markovian effects), where $\varepsilon  = 0.0001{\omega _n}$, ${q} = 0.0002{\omega _n}$, and $\omega = {\omega _n}$. For the form of Gaussian profile, we select the same parameters as the first case (the form of damped oscillation) to obtain $ \left| {\phi ({\lambda _1},\omega )} \right| \approx 0 $ when $\lambda  = {\omega _n}$ and $ \left| {\phi ({\lambda _1},\omega )} \right| \approx 0 $ when $\lambda  = 10 {\omega _n}$. The inhomogeneous term ${ a}_{in}^{(1)}(i{\lambda _1})$ is much smaller than ${ a}_{in}^{(1)}(\omega )$, which can be ignored for these parameters. The remaining five items (${ a}_{in}^{(2)}(i{\lambda _2})$, ${ b}_{in}^{(3)}(i{\lambda _3})$, ${ m}_{in}^{(4)}(i{\lambda _4})$, ${ n}_{in}^{(5)}(i{\lambda _5})$, and ${ n}_{in}^{(6)}(i{\lambda _6})$) have the similar discussions and conclusions. Therefore, the influences of the inhomogeneous term on systems parts are not considered in plotting later for studying the conversion efficiency.

Together with Eq.~(\ref{relation1}), the systems equations can be rewritten as
\begin{small}
\begin{equation}
\begin{aligned}
 - i\omega  a(\omega ) =&  i \Omega_a  a(\omega)-i{G_{ab}} b(\omega)- {{\tilde \kappa }_1}(\omega ){ a}_{in}^{(1)}(\omega )\\
 &-  a(\omega ){f_1}(\omega ) - {{\tilde \kappa }_2}(\omega ){ a}_{in}^{(2)}(\omega )-  a(\omega ){f_2}(\omega ),\\
- i\omega  b(\omega ) = & - i \omega_b  b(\omega)-i{G_{ab}} a(\omega)-i{\textsl{g}_{mb}} m(\omega)\\
&- {{\tilde \kappa }_3}(\omega ){ b}_{in}^{(3)}(\omega )-  b(\omega ){f_3}(\omega ),\\
- i\omega  m(\omega ) =&  - i \omega_m  m(\omega)-i{\textsl{g}_{mb}} b(\omega)-i{\textsl{g}_{mn}} n(\omega)\\
&- {{\tilde \kappa }_4}(\omega ){ m}_{in}^{(4)}(\omega )-  m(\omega ){f_4}(\omega ),\\
- i\omega  n(\omega ) =&  - i \omega_n  n(\omega)-i{\textsl{g}_{mn}} m(\omega)- {{\tilde \kappa }_5}(\omega ){ n}_{in}^{(5)}(\omega )\\
&-  n(\omega ){f_5}(\omega ) - {{\tilde \kappa }_6}(\omega ){ n}_{in}^{(6)}(\omega )-  n(\omega ){f_6}(\omega ),
\label{relation2}
\end{aligned}
\end{equation}
\end{small}which lead to
\begin{small}
\begin{equation}
\begin{aligned}
 a(\omega) = & - i{G_{ab}}{\eta _a}(\omega) b(\omega) - {{\tilde \kappa}_1}(\omega){\eta _a}(\omega){ a}_{in}^{(1)}(\omega) \\
&- {{\tilde \kappa}_2}(\omega){\eta _a}(\omega){ a}_{in}^{(2)}(\omega),\\
b(\omega) = & - i{G_{ab}}{\eta _b}(\omega) a(\omega) - i{\textsl{g}_{mb}}{\eta _b}(\omega) m(\omega) \\
&- {{\tilde \kappa}_3}(\omega){\eta _b}(\omega){ b}_{in}^{(3)}(\omega),\\
 m(\omega) = & - i{\textsl{g}_{mb}}{\eta _m}(\omega) b(\omega) - i{\textsl{g}_{mn}}{\eta _m}(\omega)n(\omega)\\
& - {{\tilde \kappa}_4}(\omega){\eta _m}(\omega){ m}_{in}^{(4)}(\omega),\\
 n(\omega) =  &- i{\textsl{g}_{mn}}{\eta _n}(\omega) m(\omega) - {{\tilde \kappa}_5}(\omega){\eta _n}(\omega){ n}_{in}^{(5)}(\omega)\\
& - {{\tilde \kappa}_6}(\omega){\eta _n}(\omega){ n}_{in}^{(6)}(\omega),
\label{eq21}
\end{aligned}
\end{equation}
\end{small}where the susceptibilities have the form
\begin{small}
\begin{equation}
\begin{aligned}
{\eta _a}(\omega) = &{[ - i(\omega + {\Omega _a}) + {f_1}(\omega) + {f_2}(\omega)]^{ - 1}},\\
{\eta _b}(\omega) =& {[ - i(\omega - {\omega_b}) + {f_3}(\omega)]^{ - 1}},\\
{\eta _m}(\omega) =& {[ - i(\omega - {\omega_m}) + {f_4}(\omega)]^{ - 1}},\\
{\eta _n}(\omega) = &{[ - i(\omega - {\omega_n}) + {f_5}(\omega) + {f_6}(\omega)]^{ - 1}}.
\label{eq22}
\end{aligned}
\end{equation}
\end{small}Moreover, the non-Markovian input-output relations given by Eq.~(\ref{nonMarkovian_input_output}) in the frequency domain becomes
\begin{small}
\begin{equation}
\begin{aligned}
{y }_{out}^{(\nu )}(\omega) + {y}_{in}^{(\nu )}(\omega) = -{y_\nu }(\omega ){\kappa _\nu}( - \omega ),
\label{abmn}
\end{aligned}
\end{equation}
\end{small}which causes
\begin{small}
\begin{equation}
\begin{aligned}
{\mu _{out}}(\omega) + {\mu _{in}}(\omega) = F \mu (\omega),
\label{eq177}
\end{aligned}
\end{equation}
\end{small}with $\mu(\omega) = {(a(\omega),b(\omega),m(\omega),n(\omega))^T}$, $ {\mu _{in}}(\omega) ={({a}_{in}^{(1)}(\omega),{a}_{in}^{(2)}(\omega),{{b}}_{in}^{(3)}(\omega),{{ m}}_{in}^{(4)}(\omega),{n}_{in}^{(5)}(\omega),{n}_{in}^{(6)}(\omega))^T}$, $ {\mu _{out}}(\omega) =({a}_{out}^{(1)}(\omega),{a}_{out}^{(2)}(\omega),{{b}}_{out}^{(3)}(\omega),{{ m}}_{out}^{(4)}(\omega),{n}_{out}^{(5)}(\omega)$, ${n}_{out}^{(6)}(\omega))^T$, and ${{ \kappa }_\nu}(\omega ) = \int_{-\infty}^ 0 {\kappa _\nu} (t'){e^{i\omega t' }}dt'$. The form of $6\times4$ matrix F is as follows from Eq.~(\ref{abmn})
\begin{small}
\begin{equation}
\begin{aligned}
F =- \left( {\begin{array}{*{20}{c}}
{{\kappa _1}( - \omega)}&0&0&0\\
{{\kappa _2}( - \omega)}&0&0&0\\
0&{{\kappa _3}( - \omega)}&0&0\\
0&0&{{\kappa _4}( - \omega)}&0\\
0&0&0&{{\kappa _5}( - \omega)}\\
0&0&0&{{\kappa _6}( - \omega)}
\end{array}} \right).
\label{eq16}
\end{aligned}
\end{equation}
\end{small}In the next step, we define the $4\times6$ matrix P and $4\times4$ matrix Q to bring Eq.~(\ref{relation2}) in the form
\begin{small}
\begin{equation}
\begin{aligned}
- i \omega \mu (\omega) = P{\mu _{in}}(\omega) + Q\mu (\omega).
\label{eq17}
\end{aligned}
\end{equation}
\end{small} The forms of matrices P and Q are as follows
\begin{small}
\begin{equation}
\begin{aligned}
P = -\left( {\begin{array}{*{20}{c}}
{{{\tilde  \kappa }_1}(\omega)}&{{{\tilde  \kappa }_2}(\omega)}&0&0&0&0\\
0&0&{{{\tilde  \kappa }_3}(\omega)}&0&0&0\\
0&0&0&{{{\tilde  \kappa }_4}(\omega)}&0&0\\
0&0&0&0&{{{\tilde  \kappa }_5}(\omega)}&{{{\tilde  \kappa }_6}(\omega)}
\end{array}} \right),
\label{eq19}
\end{aligned}
\end{equation}
\end{small}and
\begin{small}
\begin{equation}
\begin{aligned}
Q = \left( {\begin{array}{*{20}{c}}
{i{\Omega _a} - {q_a}}&{ - i{G_{ab}}}&0&0\\
{ - i{G_{ab}}}&{ - i{\omega_b} - {f_3}(\omega)}&{ - i{\textsl{g}_{mb}}}&0\\
0&{ - i{\textsl{g}_{mb}}}&{ - i{\omega_m} - {f_4}(\omega)}&{ - i{\textsl{g}_{mn}}}\\
0&0&{ - i{\textsl{g}_{mn}}}&{ - i{\omega_n} - {q_n}}
\end{array}} \right),
\label{eq18}
\end{aligned}
\end{equation}
\end{small}where ${q_a}={f_1}(\omega) +{f_2}(\omega)$ and ${q_n}={f_5}(\omega) +{f_6}(\omega)$.
\begin{figure*}[t]
\centering
\includegraphics[width=0.45\textwidth, clip]{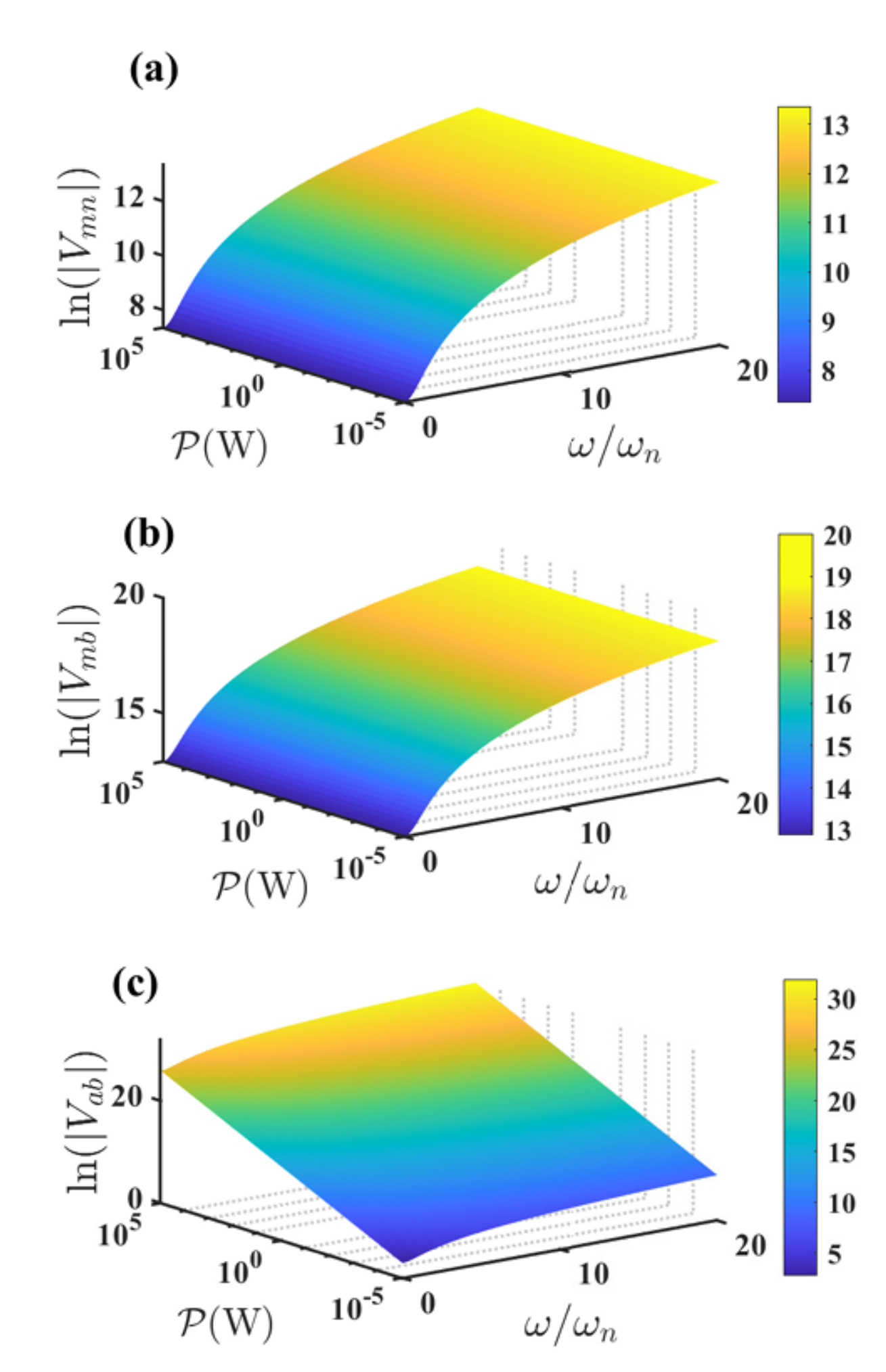}
\caption{(Color online) In non-Markovian environments, plots of the logarithms of the moduli of the magnon-microwave photon complex cooperativity ${V_{mn}}$ given by Eq.~(\ref{eq32}), the magnon-phonon complex cooperativity ${V_{mb}}$, and the optical photon-phonon complex cooperativity ${V_{ab}}$ as functions of the frequency ${\omega}$ (in units of the microwave mode frequency ${\omega_n}$) and the optical pumping power $\mathcal {P}$ (in units of W). Here, ${\lambda } = {\omega _n}$, and the other parameters are given in Table~\ref{Table1}.}
\label{3D}
\end{figure*}
\begin{figure*}[t]
\centering
\includegraphics[width=0.45\textwidth, clip]{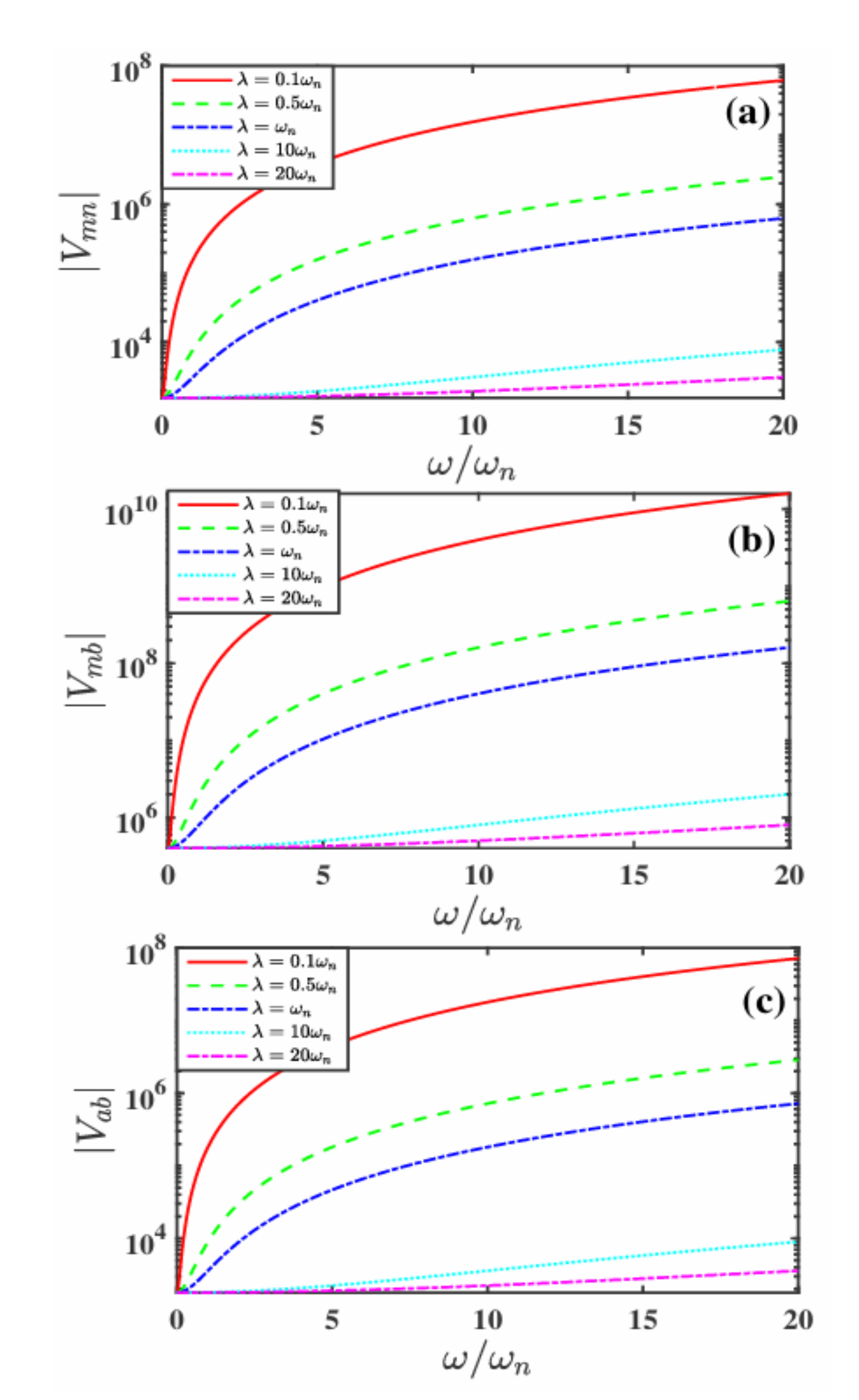}
\caption{(Color online) In a non-Markovian environment, plots of the moduli of the magnon-microwave photon complex cooperativity ${V_{mn}}$ given by Eq.~(\ref{eq32}), the magnon-phonon complex cooperativity ${V_{mb}}$, and the optical photon-phonon complex cooperativity ${V_{ab}}$ as a function of the frequency ${\omega}$ (in units of the microwave mode frequency ${\omega_n}$). The optical pumping power $\mathcal {P}=10^{-3}$ W, and the other parameters are the same as those in Fig.~\ref{3D}. }
\label{absppp}
\end{figure*}

\begin{figure*}[t]
\centering
\includegraphics[width=12cm, height=8.5cm, clip]{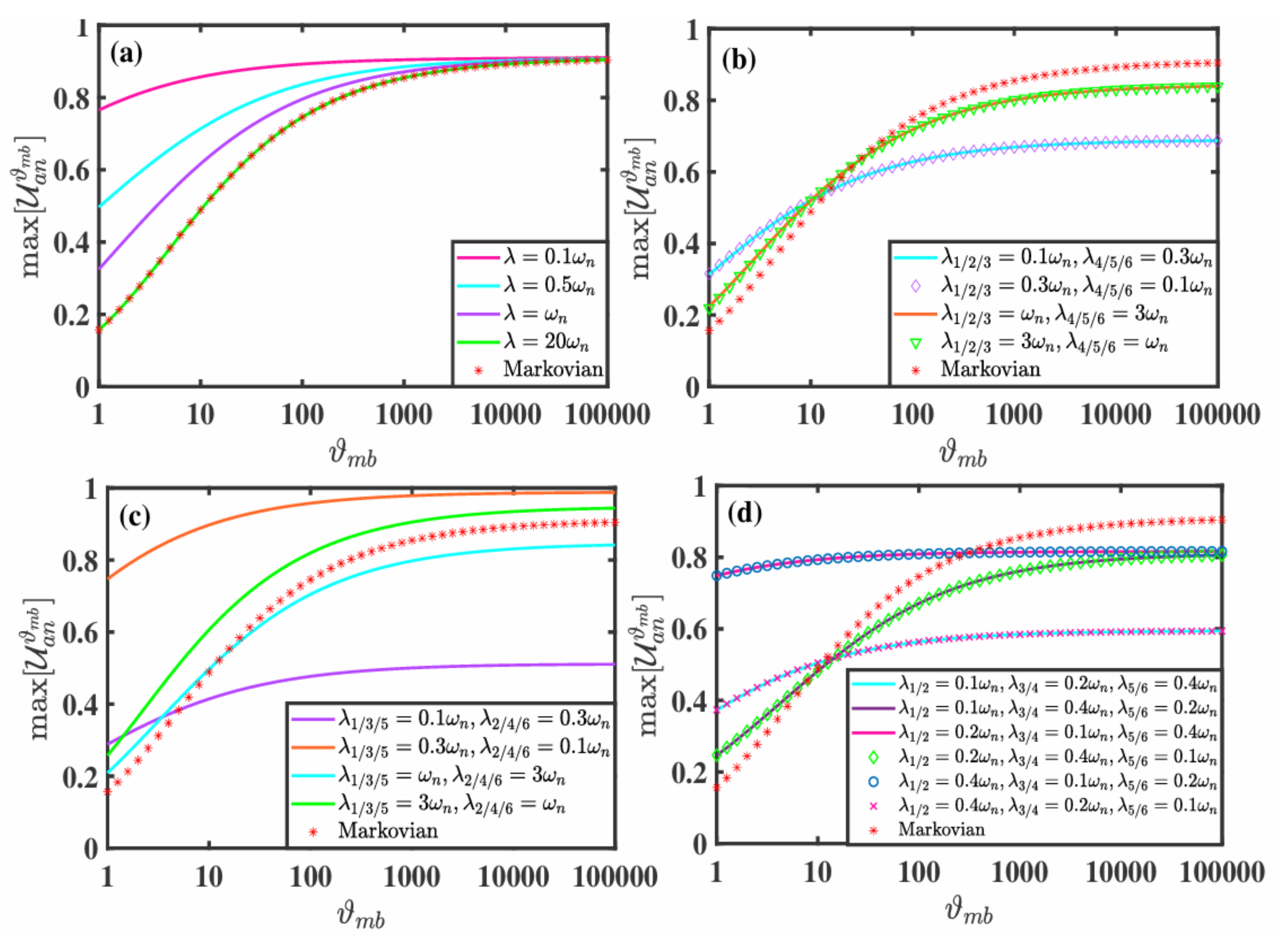}
\caption{(Color online) Maximum conversion efficiency ${\max [{\mathcal {U} _{an}^{\vartheta _{mb}}}]}$ given by Eq.~(\ref{eq33}) as a function of ${\vartheta _{mb}}$ in non-Markovian environments, where ${\vartheta _{ab}}$ and ${\vartheta _{mn}}$ fulfil Eq.~(\ref{eq351}). The cases of Markovian regimes correspond to Eq.~(\ref{cc8}). More details can be found in Appendix \ref{C}. The other parameters chosen are the same as those in Table~\ref{Table1}.}
\label{FIG3B}
\end{figure*}
We compute the scattering matrix $M(\omega)$ that relates input and output in frequency space via ${\mu _{out}}(\omega) = M(\omega){\mu _{in}}(\omega)$. We obtain the scattering matrix as a function of frequency $\omega$ referenced to a red sideband tone via
\begin{small}
\begin{equation}
\begin{aligned}
M(\omega) = F{[ - i\omega{I_4} - Q]^{ - 1}}P - {I_6},
\label{eq23}
\end{aligned}
\end{equation}
\end{small}where ${I_N}$ is the $N \times N$ identity matrix. The conversion efficiency from optics to microwave is derived as ${\mathcal {U} _{an}}(\omega) = {\left| {{M_{1,5}}(\omega)} \right|^2}$ and explicitly reads
\begin{small}
\begin{equation}
\begin{aligned}
{\mathcal {U} _{an}}(\omega) = {\left| {\frac{{{\kappa_1}( - \omega){{\tilde \kappa}_5}(\omega)\mathcal {D}{G_{ab}}{\textsl{g}_{mb}}{\textsl{g}_{mn}}}}{{(1 + \mathcal {A})(1 +\mathcal {B} ) + \mathcal {C}}}} \right|^2},
\label{eq25}
\end{aligned}
\end{equation}
\end{small}with $\mathcal {A}= \textsl{g}_{mn}^2{\eta _m}(\omega){\eta _n}(\omega), \mathcal {B}=G_{ab}^2{\eta _a}(\omega){\eta _b}(\omega), \mathcal {C}=\textsl{g}_{mb}^2{\eta _m}(\omega){\eta _b}(\omega)$, and $\mathcal {D}= {\eta _a}(\omega){\eta _b}(\omega){\eta _m}(\omega){\eta _n}(\omega)$.

Next, we define the photon-phonon complex cooperativity ${V_{ab}}$, the magnon-phonon complex cooperativity ${V_{mb}}$, and the magnon-microwave complex cooperativity ${V_{mn}}$
\begin{small}
\begin{equation}
\begin{aligned}
{V_{ab}} = {\vartheta _{ab}}{\upsilon _{ab}},\,\,\,\,\,
{V_{mb}} = {\vartheta _{mb}}{\upsilon _{mb}},\,\,\,\,\,
{V_{mn}} = {\vartheta _{mn}}{\upsilon _{mn}},\,\,\,\,
\label{eq32}
\end{aligned}
\end{equation}
\end{small}where the real cooperativities
\begin{small}
\begin{equation}
\begin{aligned}
{\vartheta _{ab}} = \frac{{4G_{ab}^2}}{{({\Gamma _1} + {\Gamma _2}){\Gamma _3}}},\,\,
{\vartheta _{mb}} = \frac{{4\textsl{g}_{mb}^2}}{{{\Gamma _3}{\Gamma _4}}},\,\,
{\vartheta _{mn}} = \frac{{4\textsl{g}_{mn}^2}}{{{\Gamma _4}({\Gamma _5} + {\Gamma _6})}},\,\,
\label{eq27}
\end{aligned}
\end{equation}
\end{small}and the influencing factors of non-Markovian effects on transport effects
\begin{small}
\begin{equation}
\begin{aligned}
{\upsilon _{ab}} =& \frac{{({\Gamma _1} + {\Gamma _2}){\Gamma _3}}}{4}/[(\frac{{{\Gamma _1}{\lambda _1}}}{{2({\lambda _1} - i\omega)}} + \frac{{{\Gamma _2}{\lambda _2}}}{{2({\lambda _2} - i\omega)}})\frac{{{\Gamma _3}{\lambda _3}}}{{2({\lambda _3} - i\omega)}}],\\
{\upsilon _{mb}} =& \frac{{({\lambda _3} - i\omega)({\lambda _4} - i\omega)}}{{{\lambda _3}{\lambda _4}}},\\
{\upsilon _{mn}} =& \frac{{{\Gamma _4}({\Gamma _5} + {\Gamma _6})}}{4}/[\frac{{{\Gamma _4}{\lambda _4}}}{{2({\lambda _4} - i\omega)}}(\frac{{{\Gamma _5}{\lambda _5}}}{{2({\lambda _5} - i\omega)}} + \frac{{{\Gamma _6}{\lambda _6}}}{{2({\lambda _6} - i\omega)}})].
\label{eq35}
\end{aligned}
\end{equation}
\end{small}

Figure \ref{3D} contains three three-dimensional surface plots, showing the logarithms of the moduli of the magnon-microwave photon complex cooperativity ${{{V}}_{mn}}$, the magnon-phonon complex cooperativity ${{{V}}_{mb}}$, and the optical photon-phonon complex cooperativity ${{{V}}_{ab}}$ given by Eq.~(\ref{eq32}) as functions of the frequency ${\omega}$ (in units of the microwave mode frequency ${\omega_n}$) and the optical pumping power $\mathcal {P}$ (in units of W). At this time, the spectral width of the non-Markovian environments ${\lambda _\nu } = {\omega _n}$, and in the subsequent discussion, we set $\lambda  = {\lambda _\nu}$. In Fig.~\ref{3D} (a), the value range of $\ln(|{{{V}}_{mn}}|)$ is from 8 to 13. The surface shows that $\ln(|{{V}_{mn}}|)$ increases as $\omega$ increases, and the growth trends are slightly different under different values of $\mathcal{P}$. In Fig.~\ref{3D} (b), the value range of $\ln(|{{{V}}_{mb}}|)$ is from 13 to 20. $\ln(|{{{V}}_{mb}}|)$ also increases as $\omega$ increases, and the overall change trend is similar to that in Fig.~\ref{3D} (a), but the value range of $\ln(|{{{V}}_{mb}}|)$ is different. In Fig.~\ref{3D} (c), the value range of $\ln(|{{{V}}_{ab}}|)$ is from 5 to 30. $\ln(|{{{V}}_{ab}}|)$ is jointly affected by $\mathcal{P}$ and $\omega$, and the shape of the surface is different from those in Fig.~\ref{3D} (a) and (b). It can be seen that the growth of $\ln(|{{{V}}_{ab}}|)$ is more obviously affected by $\mathcal{P}$.

Figure \ref{absppp} shows the plots of the moduli of the magnon-microwave photon complex cooperativity ${{{V}}_{mn}}$, the magnon-phonon complex cooperativity ${{{V}}_{mb}}$, and the optical photon-phonon complex cooperativity ${{{V}}_{ab}}$ given by Eq.~(\ref{eq32}) as functions of the frequency ${\omega}$ (in units of the microwave mode frequency ${\omega_n}$). At this time, the values of the spectral width ${\lambda }$ of the non-Markovian environment are given in the legend. The optical pumping power $\mathcal{P}$ is fixed at $10^{-3}$ W. By comparing Fig.~\ref{absppp} (a), (b) and (c), we find that the smaller the value of the environmental spectral width ${\lambda}$, that is, the stronger the non-Markovian effect, the larger the values of the moduli of the magnon-microwave photon complex cooperativity ${{{V}}_{mn}}$, the magnon-phonon complex cooperativity ${{{V}}_{mb}}$, and the optical photon-phonon complex cooperativity ${{{V}}_{ab}}$. This shows that the non-Markovian effect plays a positive role here, which can significantly increase the complex cooperativities, and further affect the conversion efficiency and bandwidth of optical and microwave photons.

In the case that all modes are on resonance, i.e., $ {\omega} = - {\Omega _a} = {\omega_b} = {\omega_m} = {\omega_n}$, the optical mode is driven by a red detuning. We can rewrite Eq.~(\ref{eq25}) in terms of the complex cooperativities ${V_{ab}}$, ${V_{mb}}$, and ${V_{mn}}$ (being necessary criterion for an efficient energy transfer)
\begin{small}
\begin{equation}
\begin{aligned}
{\mathcal {U} _{an}}(\omega) =\frac{{\mathcal {E}\mathcal {F}{V_{ab}}{V_{mb}}{V_{mn}}}}{{{{[(1 + {V_{ab}})(1 + {V_{mn}}) + {V_{mb}}]}^2}}},
\label{eq33}
\end{aligned}
\end{equation}
\end{small}with $\mathcal {E}=\frac{{\kappa _1^2(-\omega )}}{{[{f_1}(\omega ) + {f_2}(\omega )]}}$ and $\mathcal {F}= \frac{{\tilde \kappa _5^2(\omega )}}{{[{f_5}(\omega ) + {f_6}(\omega )]}}$.

Thus, we take into account a configuration where the free variables are the magnon-microwave real cooperativity ${\vartheta _{mn}}$ and the photon-phonon real cooperativity ${\vartheta _{ab}}$, which leads to the constraint
\begin{small}
\begin{equation}
\begin{aligned}
{\vartheta _{ab}} &\equiv \frac{{{i}\sqrt {1 + {\upsilon _{mb}}{\vartheta _{mb}} + {\upsilon _{mn}}{\vartheta _{mn}}} \sqrt {1 + \upsilon _{mb}^ * {\vartheta _{mb}} + \upsilon _{mn}^ * {\vartheta _{mn}}} }}{{\sqrt { - {\upsilon _{ab}}\upsilon _{ab}^ * \left( {1 + {\upsilon _{mn}}{\vartheta _{mn}}} \right)\left( {1 + \upsilon _{mn}^ * {\vartheta _{mn}}} \right)} }},\\
{\vartheta _{mn}} &\equiv \frac{{{i}\sqrt {1 + {\upsilon _{ab}}{\vartheta _{ab}} + {\upsilon _{mb}}{\vartheta _{mb}}} \sqrt {1 + \upsilon _{ab}^ * {\vartheta _{ab}} + \upsilon _{mb}^ * {\vartheta _{mb}}} }}{{\sqrt { - {\upsilon _{mn}}\upsilon _{mn}^ * \left( {1 + {\upsilon _{ab}}{\vartheta _{ab}}} \right)\left( {1 + \upsilon _{ab}^ * {\vartheta _{ab}}} \right)} }},
\label{eq35}
\end{aligned}
\end{equation}
\end{small}by fixing ${\vartheta _{mb}}$ and setting the partial derivatives of Eq.~(\ref{eq33}) with respect to ${\vartheta _{mn}}$ and ${\vartheta _{ab}}$ to zero for the real cooperativities. In this case, we have the maximum efficiency at
\begin{small}
\begin{equation}
\begin{aligned}
{\vartheta _{ab}} = {\vartheta _{mn}},
\label{eq351}
\end{aligned}
\end{equation}
\end{small}so that we reach a quartic equation about ${\vartheta _{ab}}$
\begin{small}
\begin{equation}
\begin{aligned}
\mathcal{G}\vartheta _{ab}^4 + \mathcal{H}\vartheta _{ab}^3 + \mathcal{I}\vartheta _{ab}^2 - \mathcal{J}{\vartheta _{ab}} - \mathcal{K} = 0,
\label{eqq351}
\end{aligned}
\end{equation}
\end{small}where $\mathcal{G} = {\upsilon _{ab}}\upsilon _{ab}^*{\upsilon _{mn}}\upsilon _{mn}^*, \mathcal{H} = {\upsilon _{mn}}\upsilon _{mn}^*({\upsilon _{ab}} + \upsilon _{ab}^*), \mathcal{I} = {\upsilon _{mn}}\upsilon _{mn}^* - {\upsilon _{ab}}\upsilon _{ab}^*, \mathcal{J} = ({\upsilon _{mb}}\upsilon _{ab}^* + {\upsilon _{ab}}\upsilon _{mb}^*){\vartheta _{mb}} + ({\upsilon _{ab}} + \upsilon _{ab}^*)$, and $\mathcal{K} = {\upsilon _{mb}}\upsilon _{mb}^*\vartheta _{mb}^2 + ({\upsilon _{mb}} + \upsilon _{mb}^*){\vartheta _{mb}} + 1$.

We know that under the influence of non-Markovian effects, the conversion efficiency close to unit can be achieved without the need for large values of ${\vartheta _{ab}}$ and ${\vartheta _{mn}}$. Due to the different values of $\lambda$, the maximum conversion efficiency also varies in Fig.~\ref{FIG3B}. We take six $\lambda$ values as equal, as shown in Fig.~\ref{FIG3B}(a), the smaller the value of $\lambda$, the stronger the non-Markovian effects, the higher the maximum conversion efficiency in regions with smaller orders of magnitude in ${\vartheta _{mb}}$. Eventually, each maximum conversion efficiency tends to 0.9 at  ${\vartheta _{mb}} = 4 \times {10^5}$. When $\lambda  = 20{\omega _n}$, the curve coincides with the Markovian curve, which corresponds to Eq.~(\ref{cc8}). More details can be found in Appendix \ref{C}. (i) If the values of $\lambda$ are different, setting the parameters of both has a triple relationship with each other when ${\lambda _1} = {\lambda _2} = {\lambda _3}$ and ${\lambda _4} = {\lambda _5} = {\lambda _6}$ as shown in Fig.~\ref{FIG3B}(b).
We find that regardless of the exchange of values between ${\lambda _{1/2/3}}$ and ${\lambda _{4/5/6}}$, there is no influence on the maximum conversion efficiency. As the values of ${\lambda _{1/2/3}}$ and ${\lambda _{4/5/6}}$ increasing, their maximum conversion efficiency ultimately increasing, but do not exceed that of Markovian regimes. (ii) As shown in Fig.~\ref{FIG3B}(c), the parameters chosen are still in a triple relationship with each other when ${\lambda _1} = {\lambda _3} = {\lambda _5}$ and ${\lambda _2} = {\lambda _4} = {\lambda _6}$. The results at this point are completely different from (i). When the value of ${\lambda _{1/3/5}}$ is larger than that of ${\lambda _{2/4/6}}$, the corresponding maximum conversion efficiencies are larger and higher than those of Markovian regimes. On the contrary, when the value of ${\lambda _{1/3/5}}$ is smaller than that of ${\lambda _{2/4/6}}$, the corresponding maximum conversion efficiencies are smaller and lower than those of Markovian regimes. The enhancement of maximum conversion efficiency can be explained by the excitation backflowing originating from the non-Markovian effects of the environments to systems. This indicates that the feedback effects of the non-Markovian environment acting on the systems dynamics can enhance the maximum conversion efficiency. (iii) When the three parameters are in a double relationship with each other in Fig.~\ref{FIG3B}(d), where ${\lambda _1} = {\lambda _2}$, ${\lambda _3} = {\lambda _4}$, and ${\lambda _5} = {\lambda _6}$. It is interesting that the two curves exchanging the values of ${\lambda _{1/2}}$ and ${\lambda _{5/6}}$ overlap, but in the end, the maximum conversion efficiencies are not as high as those of Markovian regimes. Taking other possible values for environmental spectral widths can also be studied, but we do not intend to go into details here. Readers interested in this point can give it a try.

\section{the conversion efficiency and the bandwidths with non-Markovian effects}
\begin{figure}[t]
\centering
\includegraphics[width=7cm, height=9cm, clip]{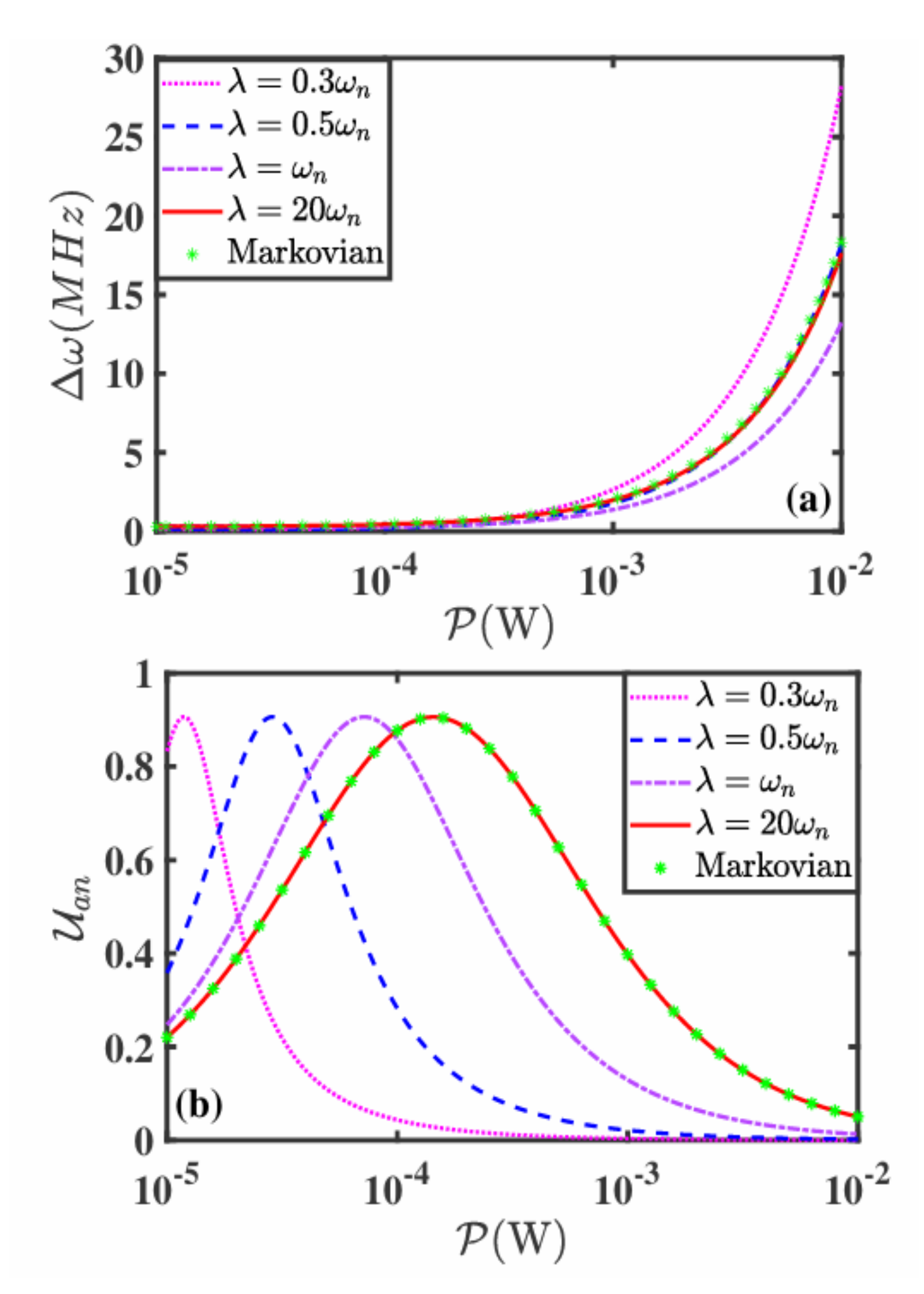}
\caption{(Color online) (a) Bandwidth ${\Delta {\omega}}$ given by Eq.~(\ref{band}) as a function of optical pump power $\mathcal {P}$ with non-Markovian effects. The conversion bandwidth of Markovian regimes corresponds to Eq.~(\ref{c8}), which can be found in Appendix \ref{C}. (b) Conversion efficiency ${\mathcal {U} _{an}}(\omega)$ given by Eq.~(\ref{eq33}) as a function of optical pump power $\mathcal {P}$ in non-Markovian environments. Eq.~(\ref{ceq33}) defines the conversion efficiency of Markovian regimes in Appendix~\ref{C}. We set $ {\omega} = - {\Omega _a} = {\omega_b} = {\omega_m} = {\omega_n}$ and $\lambda  = {\lambda _\nu}$. The other parameters chosen are the same as those in Table~\ref{Table1}.}
\label{bandwidthandeff}
\end{figure}
We find the value $\Delta {\omega _{1/2}}$ where the efficiency drops off by 50 $\%$ compared to the values at ${\omega _{\max }}$. The difference between $ {\omega _{\max }}$ and ${\Delta {\omega _{1/2}}}$ gives half of the conversion bandwidth. The conversion bandwidth is therefore defined as ${\Delta {\omega}} =2 {\Delta {\omega _{1/2}}} $. The half width at full maximum $\Delta {\omega _{1/2}}$ of a peak in the transmission spectrum at frequency $ {\omega _{\max }}$ is defined by
\begin{small}
\begin{equation}
\begin{aligned}
{\mathcal {U} _{an}}({\omega _{\max }} \pm \Delta {\omega _{1/2}}) =\frac{{\mathcal {U} _{an}}( {\omega _{\max }})}{2}.
\label{band}
\end{aligned}
\end{equation}
\end{small}The frequency $ {\omega _{\max }}$ at maximum conversion efficiency is actually equal to $ {\omega _n}$. In Fig.~\ref{bandwidthandeff}(a), our analysis shows that it can return to the similar results as Markovian environments when ${\lambda} = 20 {\omega _n}$. The conversion bandwidth and the conversion efficiency of Markovian regimes correspond to Eq.~(\ref{c8}) and Eq.~(\ref{ceq33}), respectively, which can be found in Appendix \ref{C}. Decreasing the environmental spectral widths from $20 {\omega _n}$ to ${\omega _n}$, the conversion bandwidth in Eq.~(\ref{band}) decreases from 18 MHz to 13 MHz at the optical pump power $\mathcal {P} = 0.01$ W. The conversion bandwidth is also similar to the Markovian case when ${\lambda} = 0.5 {\omega _n}$. However, we discover that the conversion bandwidth increases to 28 MHz at the optical pump power $\mathcal {P} = 0.01$ W, where ${\lambda} = 0.3 {\omega _n}$. This indicates that when $\lambda  < 0.5 {\omega _n} $, the non-Markovian effects can play very positive roles in increasing the conversion bandwidths, which originates from the excitation backflowing obtained by the interaction between systems and environments. The increase in conversion bandwidth leads to a decrease in conversion efficiency in Eq.~(\ref{eq33}), as shown in Fig.~\ref{bandwidthandeff}(b). The peak of conversion efficiency remains 0.9, no matter how ${\lambda}$ is adjusted. As the value of ${\lambda}$ becomes smaller, the non-Markovian effects gradually strengthen, which means that we do not need such a large optical pump power to achieve the same maximum conversion efficiency as in the Markovian environmental regimes.
In this regime, we achieve a very large bandwidth at the cost of a decreased conversion efficiency.
\begin{figure*}[t]
\centering
\includegraphics[width=10cm, height=4cm, clip]{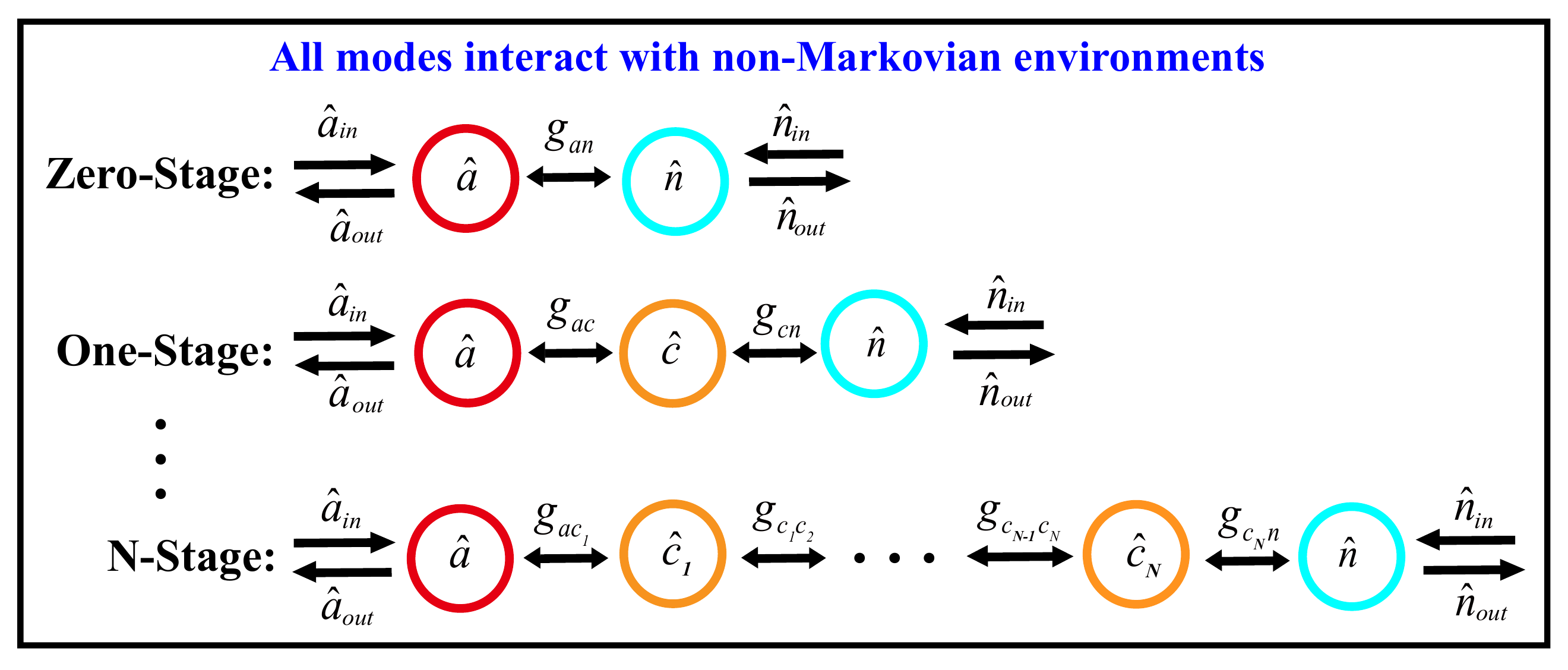}
\caption{(Color online) Illustration of the conversion pathway between optical photon mode ($ \hat a $) and microwave photon mode ($ \hat n $) in non-Markovian environments. ${\hat a_{in}}$, ${\hat a_{out}}$, ${\hat n_{in}}$ and ${\hat n_{out}}$ are the input and output operators for optical photon mode and microwave photon mode, respectively. ${\textsl{g}_{ij}}$ denotes the interacting strength between each mode. In general, the conversion process can involve $N$ intermediate modes (${\hat c_j}$ with $j = 0,1, \cdots ,N + 1,N$) as the figure presented.}
\label{Nstage}
\end{figure*}

\section{CONCLUSIONS}
In summary, we have studied a two-stage frequency conversion between optical and microwave photons based on magnetic and mechanical excitations in non-Markovian environments. By adjusting the complex cooperativity, the conversion efficiency close to unity can be achieved. In non-Markovian regimes, the environmental spectral widths enhanced conversion efficiency and bandwidths due to excitation backflowing from interactions between the systems and environments. A large order of magnitude of optical pump power can increase conversion bandwidth but decrease conversion efficiency. Conversely, decreasing spectral width can maintain maximum efficiency without requiring high optical pump power in non-Markovian environments. Additionally, weaker magnon-microwave interaction led to a larger conversion bandwidth. Our results did not take into account any specific geometry and can be used as a starting point for designing a magnomechanical-based microwave-to-optics converter in non-Markovian environments.

The approach suggested in this work widens a variety of potential applications for quantum communication and information \cite{Salathe90340112018}, which helps clarify the relationship between non-Markovianity and frequency conversion. Our results might also be extended to wide class of open quantum systems with (1) second-order nonlinear materials $G({{\hat b}^2}{{\hat c}^\dag } + \hat c{{\hat b}^{\dag 2}})$ \cite{Majumdar872353192013,Shen900238492014}, (2) third-order Kerr nonlinearity $U{{\hat b}^{\dag 2}}{{\hat b}^2}$ \cite{Ferretti850333032012,Shen910638082015}, (3) Jaynes-Cummings Hamiltonian $\sum\nolimits_j {{V_j}({{\hat \sigma} _ - }{\hat b}_j^\dag  + {{\hat b}_j}{{\hat \sigma} _ + })} $ \cite{Shen990321012019,Shen930121072016}, or Rabi models $\sum\nolimits_j {{G_j}{{\hat \sigma} _x}({\hat b}_j^\dag  + {{\hat b}_j})} $ \cite{Shi1201536022018,Lo980638072018}, and (4) quadratic optomechanical couplings $ U{\hat c^\dag }\hat c(\hat b{\rm{ + }}{\hat b^\dag })^2 $ \cite{Aspelmeyer8613912014,Xuereb870238302013} interacting with non-Markovian reservoirs, which deserves future studies. Moreover, the frequency conversion process between optical photons and microwave photons can involve $N$ intermediate media \cite{Han810502021} in non-Markovian environments as shown in Fig.~\ref{Nstage}. We can also try to extend the two-stage converter to $N$-stage converter to explore higher conversion efficiency and wider conversion bandwidth with non-Markovian effects.

The quantum transduction \cite{Lauk50205012020} of microwave and optical fields is currently a very active area of research with non-Markovian effects. There has been a lot of progress in a relatively short amount of time, with the good systems operating already at the few-photon level and relatively high efficiency. However, this challenging endeavor is far from being completed, with a lot of interesting physics still lying ahead of us. These developments are driven by our rapidly evolving abilities to experimentally manipulate and control quantum dynamics in open quantum systems. Quantum communication and quantum information processing may before long benefit from such quantum technologies in non-Markovian environments.
\section*{ACKNOWLEDGMENTS}
This work was supported by National Natural Science Foundation of China under Grants No.~12274064, and Scientific Research Project for Department of Education of Jilin Province under Grant No.~JJKH20241410KJ. 
During the writing of this paper, Jia Tang would like to thank all the valuable feedback and suggestions that have prompted me to continuously improve, supplement, and enhance the content and quality of the paper.
\appendix
\section{Derivation of the effective linearized Hamiltonian in Eq.~(\ref{eq9})}
\label{A}
Here, we give a detailed derivation of the effective linearized Hamiltonian (\ref{eq9}) with some approximations \cite{Xie250730092023,Potts130640012020,Potts110310532021}. Using the Hamiltonian in Eq.~(\ref{eq1}) as a starting point, we transform the Hamiltonian to a frame rotating at the drive frequency ${\omega_D}$ by applying the unitary transformation. By applying a rotating-wave approximation, we disregard terms in form ${\hat m}{\hat b}$, ${\hat m^{\dag }}{\hat b^{\dag}}$, ${\hat m}{\hat n}$, ${\hat m^{\dag}}{\hat n^{\dag}}$ and get
\begin{small}
\begin{equation}
\begin{aligned}
{{\hat H'}_1}/\hbar= & - {\Omega_a}{{\hat a}^\dag }\hat a + {\omega_b}{{\hat b}^\dag }\hat b + {\omega_m}{{\hat m}^\dag }\hat m + {\omega_n}{{\hat n}^\dag }\hat n\\
&+ {\textsl{g}_{ab}}{{\hat a}^\dag }\hat a({{\hat b}^\dag } + \hat b) + {\textsl{g}_{mb}}({{\hat m}^\dag }\hat b+ \hat m {{\hat b}^\dag })\\
              &+ {\textsl{g}_{mn}}({{\hat m}^\dag }\hat n+ \hat m {{\hat n}^\dag }) + i\mathbbm{a} (\hat a - {\hat a^\dag })\\
              &+\sum\limits_{\nu,k} {c_k^{(\nu)} } \hat d_k^{{(\nu)} \dag }\hat d_k^{(\nu)} + \sum\limits_{\nu,k} {(x_k^{(\nu)} } { \hat y_\nu }\hat d_k^{{(\nu)} \dag } + x_k^{{(\nu)} *}\hat y_\nu ^\dag \hat d_k^{(\nu)}).
\label{a2}
\end{aligned}
\end{equation}
\end{small}Substituting the Hamiltonian of Eq.~(\ref{a2}) into the Heisenberg equation $\dot {\hat A}(t) =  - i[\hat A(t),{{\hat H'}_1}(t)]$ yields
\begin{small}
\begin{equation}
\begin{aligned}
{\dot {\hat a}} = &i{\Omega _a} \hat a - i{\textsl{g}_{ab}}(\hat b + {{\hat b}^\dag })\hat a - \mathbbm{a} - i\sum\limits_k x_k^{{(1)}*}{\hat d}_k^{(1)}\\
& - i \sum\limits_k x_k^{{(2)}*}{\hat d}_k^{(2)},\\
{\dot {\hat b}} =&  - i{\omega_b} \hat b - i{\textsl{g}_{ab}} {{\hat a}^\dag }\hat a - i{\textsl{g}_{mb}}\hat m - i \sum\limits_k x_k^{{(3)}*}{\hat d}_k^{(3)},\\
{\dot {\hat m}} = & - i{\omega_m} \hat m - i{\textsl{g}_{mb}}\hat b- i{\textsl{g}_{mn}}\hat n - i \sum\limits_k x_k^{{(4)}*}{\hat d}_k^{(4)},\\
{\dot {\hat n}} = & - i{\omega_n} \hat n - i{\textsl{g}_{mn}}\hat m - i \sum\limits_k x_k^{{(5)}*}{\hat d}_k^{(5)} - i \sum\limits_k x_k^{{(6)}*}{\hat d}_k^{(6)},\\
\dot {\hat d}_k^{(\nu)}=&  - i  {{c_k^{(\nu)}}} {{\hat d}_k^{(\nu)}} - i {{x_k^{(\nu)}}} {{\hat y}_\nu }.
\label{a3}
\end{aligned}
\end{equation}
\end{small}Employing standard quantum optics procedures, we write the operator in the form of the sum of steady-state values and quantum fluctuation operators as follows $\hat a = \mathbbm{a}  + \delta \hat a$, $\hat b = \mathbbm{b} + \delta \hat b$, $\hat m = \mathbbm{m}  + \delta \hat m$, $\hat n = \mathbbm{n}  + \delta \hat n$, ${\hat d}_k^{(\nu)} = \mathbbm{d}_k^{(\nu)}  + \delta {\hat d}_k^{(\nu)}$. Substituting the above decompositions into Eq.~(\ref{a3}), we obtain
\begin{small}
\begin{equation}
\begin{aligned}
{\dot {\mathbbm{a}}} = & i{\Omega _a} \mathbbm{a} - i{\textsl{g}_{ab}}(\mathbbm{b}+{\mathbbm{b}^*})\mathbbm{a} - {\mathbbm{a}} - i \sum\limits_k x_k^{{(1)}*} \mathbbm{d}_k^{{(1)}}\\& - i \sum\limits_k x_k^{{(2)}* \mathbbm{d}_k^{{(2)}}}=0,\\
{\dot {\mathbbm{b}}} = & - i{\omega_b} \mathbbm{b} - i{\textsl{g}_{ab}} {\mathbbm{a}^* }\mathbbm{a} - i{\textsl{g}_{mb}}\mathbbm{m} - i \sum\limits_k x_k^{{(3)}*} \mathbbm{d}_k^{{(3)}} =0,\\
{\dot {\mathbbm{m}}} = & - i{\omega_m} \mathbbm{m} - i{\textsl{g}_{mb}}\mathbbm{b}- i{\textsl{g}_{mn}}\mathbbm{n} - i \sum\limits_k x_k^{{(4)}*} \mathbbm{d}_k^{{(4)}} =0,\\
{\dot {\mathbbm{n}}} = & - i{\omega_n} \mathbbm{n} - i{\textsl{g}_{mn}}\mathbbm{m}- i \sum\limits_k x_k^{{(5)}*} \mathbbm{d}_k^{{(5)}} - i \sum\limits_k x_k^{{(6)}*} \mathbbm{d}_k^{{(6)}}\\&=0 ,\\
{\dot {\mathbbm{d}}}_k^{(\nu)}= & - i  {{c_k^{(\nu)}}} \mathbbm{d}_k^{(\nu)} - i  {{x_k^{(\nu)}}} \mathbbm{y}_\nu =0,
\label{a4}
\end{aligned}
\end{equation}
\end{small}where $ {{\mathbbm{y}}_1} = {{\mathbbm{y}}_2} =  \mathbbm{a}$, ${\mathbbm{y}_3} =\mathbbm{b}$, ${{\mathbbm{y}}_4} = \mathbbm{m}$, and ${{\mathbbm{y}}_5} = {{\mathbbm{y}}_6} = \mathbbm{n}$.

In this case, the quantum fluctuation operators satisfy
\begin{small}
\begin{equation}
\begin{aligned}
{\delta {\dot {\hat a}}} = &   i{\Omega _a} {\delta {\hat a}} - i{G_{ab}}({\delta {\hat b}} + {\delta{{\hat b}^\dag }})- i \sum\limits_k x_k^{{(1)}*}\delta \hat d_k^{(1)} - i \sum\limits_k x_k^{{(2)}*}\delta \hat d_k^{(2)},\\
{\delta {\dot {\hat b}}} = & - i{\omega_b} {\delta \hat b} - i{G_{ab}} ({{\delta {\hat a}}^\dag }+{\delta {\hat a}}) - i{\textsl{g}_{mb}}{\delta {\hat m}} - i \sum\limits_k x_k^{{(3)}*}\delta \hat d_k^{(3)},\\
{\delta {\dot {\hat m}}} = & - i{\omega_m} {\delta \hat m} - i{\textsl{g}_{mb}}{\delta {\hat b}}- i{\textsl{g}_{mn}}{\delta{\hat n}} - i \sum\limits_k x_k^{{(4)}*}\delta \hat d_k^{(4)},\\
{\delta {\dot {\hat n}}} = & - i{\omega_n} {\delta \hat n} - i{\textsl{g}_{mn}}{\delta {\hat m}} - i \sum\limits_k x_k^{{(5)}*}\delta \hat d_k^{(5)} - i \sum\limits_k x_k^{{(6)}*}\delta \hat d_k^{(6)},\\
{\delta {{\dot {\hat d}}_k^{(\nu)}}}= &- i {{c_k^{(\nu)}}} {\delta {{\hat d}_k^{(\nu)}}} - i {{x_k^{(\nu)}}} {\delta {\hat y_\nu}},
\label{a6}
\end{aligned}
\end{equation}
\end{small}which lead to the linearized Hamiltonian
\begin{small}
\begin{equation}
\begin{aligned}
{{\hat H}^{lin}} /\hbar=& - {\Omega _a}{\delta{{\hat a}^\dag }}{\delta {\hat a}} + {\omega_b}{\delta{{\hat b}^\dag }}{\delta {\hat b}} + {\omega_m}{\delta{{\hat m}^\dag }}{\delta {\hat m}} + {\omega_n}{\delta{{\hat n}^\dag }}{\delta {\hat n}}\\
            &+ {G_{ab}}({\delta{{\hat a}^\dag }}+{\delta {\hat a}})({\delta {\hat b}}+ {\delta{{\hat b}^\dag }}) + {\textsl{g}_{mb}}({\delta{{\hat m}^\dag }}{\delta {\hat b}}+ {\delta {\hat m}} {\delta{{\hat b}^\dag }})\\
              &+ {\textsl{g}_{mn}}({\delta{{\hat m}^\dag }}{\delta {\hat n}}+ {\delta{\hat m}} {\delta{{\hat n}^\dag }})\\
             &+\sum\limits_{\nu,k} {c_k^{(\nu)} } \delta \hat d_k^{{(\nu)} \dag }\delta \hat d_k^{(\nu)}\\& + \sum\limits_{\nu,k} {(x_k^{(\nu)} } \delta {\hat y_\nu }\delta \hat d_k^{{(\nu)} \dag } + x_k^{{(\nu)} *}\delta \hat y_\nu ^\dag \delta \hat d_k^{(\nu)} ).
\label{a7}
\end{aligned}
\end{equation}
\end{small}For a red-detuned drive ${\Omega _a} \approx  - {\omega_b}$, the terms $\delta {\hat a}\delta {\hat b}$ and $\delta {\hat a^{\dag }}\delta {\hat b^{\dag }}$ can also be neglected under the rotating-wave approximation. This causes the Hamiltonian (\ref{eq9}), where we have re-defined $\delta \hat a \to \hat a$, $\delta \hat b \to \hat b$, $\delta \hat m \to \hat m$, $\delta \hat n \to \hat n$, $\delta {{\hat d}_k^{(\nu)}} \to {{\hat d}_k^{(\nu)}}$.
\section{Derivation of Lorentzian resonance spectrum and conversion efficiency and bandwidth in the Markovian regimes}
\label{C}
Defining six auxiliary modes ${s_\nu}(t) = \int_0^t {d\tau  {y_\nu}(\tau )} {f_\nu}(t - \tau )$ in non-Markovian environments, Eq.~(\ref{eq13}) becomes
\begin{small}
\begin{equation}
\begin{aligned}
{\dot a}(t) = &i{\Omega _a} a(t) - i{G_{ab}} b(t) -  K_1 (t) - K_2 (t)- {s_1}(t) - {s_2}(t),\\
{\dot b}(t) = & - i{\omega_b} b(t) - i{G_{ab}} a(t) - i{\textsl{g}_{mb}} m(t) - K_3 (t) -{s_3}(t),\\
{\dot m}(t) = & - i{\omega_m} m(t) - i{\textsl{g}_{mb}} b(t) - i{\textsl{g}_{mn}} n(t) -  K_4 (t) -{s_4}(t),\\
{\dot n}(t) = & - i{\omega_n} n(t) - i{\textsl{g}_{mn}} m(t) - K_5 (t) -  K_6 (t)- {s_5}(t)\\&- {s_6}(t),\\
{\dot s_\nu}(t) = & \frac{1}{2}{\Gamma _\nu}{\lambda _\nu} {y_\nu}(t) - {\lambda _\nu}{s_\nu}(t).
\label{c1}
\end{aligned}
\end{equation}
\end{small}The differential equations of Eq.~(\ref{c1}) can be written as matrix in the form of $\dot \varpi (t) = \mathscr{A}\varpi (t) + \mathscr{B}$ with
$\varpi (t)= ( a(t), b(t), m(t), n(t), {s_1}(t), {s_2}(t), {s_3}(t), {s_4}(t)$, ${s_5}(t), {s_6}(t))^T$, where
\begin{small}
\begin{widetext}
\begin{equation}
\begin{aligned}
\mathscr{A} = \left( {\begin{array}{*{20}{c}}
{i{\Omega _a}}&{ - i{G_{ab}}}&0&0&{ - 1}&{ - 1}&0&0&0&0\\
{ - i{G_{ab}}}&{ - i{\omega _b}}&{ - i{g_{mb}}}&0&0&0&{ - 1}&0&0&0\\
0&{ - i{g_{mb}}}&{ - i{\omega _m}}&{ - i{g_{mn}}}&0&0&0&{ - 1}&0&0\\
0&0&{ - i{g_{mn}}}&{ - i{\omega _n}}&0&0&0&0&{ - 1}&{ - 1}\\
{\frac{1}{2}{\Gamma _1}{\lambda _1}}&0&0&0&{ - {\lambda _1}}&0&0&0&0&0\\
{\frac{1}{2}{\Gamma _2}{\lambda _2}}&0&0&0&0&{ - {\lambda _2}}&0&0&0&0\\
0&{\frac{1}{2}{\Gamma _3}{\lambda _3}}&0&0&0&0&{ - {\lambda _3}}&0&0&0\\
0&0&{\frac{1}{2}{\Gamma _4}{\lambda _4}}&0&0&0&0&{ - {\lambda _4}}&0&0\\
0&0&0&{\frac{1}{2}{\Gamma _5}{\lambda _5}}&0&0&0&0&{ - {\lambda _5}}&0\\
0&0&0&{\frac{1}{2}{\Gamma _6}{\lambda _6}}&0&0&0&0&0&{ - {\lambda _6}}
\end{array}} \right), \, \, \, \, \, \, \, \, \,
\mathscr{B}=\left( {\begin{array}{*{20}{c}}
{ - K_1 (t)- K_2 (t)}\\
{- K_3 (t)}\\
{-  K_4 (t)}\\
{ -K_5 (t) - K_6 (t)}\\
0\\
0\\
0\\
0\\
0\\
0
\end{array}} \right).
\label{c3}
\end{aligned}
\end{equation}
\end{widetext}
\end{small}Under the Markovian approximation, Eq.~(\ref{eq13}) reads
\begin{small}
\begin{equation}
\begin{aligned}
{\dot a}(t) = &-i(-{\Omega _a} -i \frac{{{\Gamma _1} + {\Gamma _2}}}{2})a(t) - i{G_{ab}} b(t)\nonumber \\
&+i\sqrt {{\Gamma _1}} { a}_{in}^{(1)}(t)+i \sqrt {{\Gamma _2}} { a}_{in}^{(2)}(t),\nonumber
\end{aligned}
\end{equation}
\end{small}
\begin{small}
\begin{equation}
\begin{aligned}
{\dot b}(t) = & - i({\omega_b} -i \frac{{{\Gamma _3}}}{2}) b(t) - i{G_{ab}} a(t) - i{\textsl{g}_{mb}} m(t)\\&+i\sqrt {{\Gamma _3}} {b}_{in}^{(3)}(t),\nonumber
\end{aligned}
\end{equation}
\end{small}
\begin{small}
\begin{equation}
\begin{aligned}
{\dot m}(t) = & - i({\omega_m}-i \frac{{{\Gamma _4}}}{2}) m(t) - i{\textsl{g}_{mb}} b(t) - i{\textsl{g}_{mn}} n(t) \\&+i\sqrt {{\Gamma _4}} {m}_{in}^{(4)}(t),\nonumber \\
\end{aligned}
\end{equation}
\end{small}
\begin{small}
\begin{equation}
\begin{aligned}
{\dot n}(t) = & - i({\omega_n}-i \frac{{{\Gamma _5} + {\Gamma _6}}}{2}) n(t) - i{\textsl{g}_{mn}} m(t) \\
&+i\sqrt {{\Gamma _5}} { n}_{in}^{(5)}(t)+i \sqrt {{\Gamma _6}} { n}_{in}^{(6)}(t),
\label{c4}
\end{aligned}
\end{equation}
\end{small}which can be denoted as $\dot \chi (t) = \mathscr{C}{\chi _{in}}(t)+\mathscr{D} \chi (t)$ with $\chi (t) = {( a(t), b(t), m(t), n(t))^T}$ and $ {\chi _{in}}(t) ={( { a}_{in}^{(1)}(t), { a}_{in}^{(2)}(t),{{b}}_{in}^{(3)}(t),{{m}}_{in}^{(4)}(t),{n}_{in}^{(5)}(t), { n}_{in}^{(6)}(t))^T}$, where
\begin{small}
\begin{equation}
\begin{aligned}
\mathscr{C}=\left( {\begin{array}{*{20}{c}}
{i\sqrt {{\Gamma _1}}}&{i\sqrt {{\Gamma _2}}}&0&0&0&0\\
0&0&{i\sqrt {{\Gamma _3}}}&0&0&0\\
0&0&0&{i\sqrt {{\Gamma _4}}}&0&0\\
0&0&0&0&{i\sqrt {{\Gamma _5}}}&{i\sqrt {{\Gamma _6}}}
\end{array}} \right),
\label{c6}
\end{aligned}
\end{equation}
\end{small}and
\begin{small}
\begin{equation}
\begin{aligned}
\mathscr{D} = \left( {\begin{array}{*{20}{c}}
{i{\Omega _a} - \frac{{\Gamma _a}}{2}}&{ - i{G_{ab}}}&0&0\\
{ - i{G_{ab}}}&{  - i{\omega_b} - \frac{{{\Gamma _3}}}{2}}&{ - i{g_{mb}}}&0\\
0&{ - i{g_{mb}}}&{ - i{\omega_m}-\frac{{{\Gamma _4}}}{2}}&{ - i{g_{mn}}}\\
0&0&{ - i{g_{mn}}}&{- i{\omega_n}- \frac{{\Gamma _n}}{2}}\end{array}} \right),
\label{c5}
\end{aligned}
\end{equation}
\end{small}with ${\Gamma_a} = {\Gamma_1} + {\Gamma_2}$ and ${\Gamma_n} = {\Gamma_5} + {\Gamma_6}$. Matrix $\mathscr{D} $ is a $4\times4$ matrix and decreases six dimensions compared to the $10\times10$ matrix $\mathscr{A} $ in the non-Markovian case. This is because when considering the Markovian approximation, Eq.~(\ref{c4}) no longer has integral differential terms. Therefore, there is no need to introduce auxiliary modes, where the systems differential equations are closed and solvable.

The frequency conversion efficiency from optics to microwave in Markovian regimes is
\begin{small}
\begin{equation}
\begin{aligned}
{ {\mathcal {U}} _{an}}(\omega) = {\left| {\frac{{\sqrt{{\Gamma_1}{\Gamma_5}}{\mathcal {D}}{G_{ab}}{\textsl{g}_{mb}}{\textsl{g}_{mn}}}}{{(1 + {\mathcal {A}})(1 +{\mathcal {B}} ) + {\mathcal {C}}}}} \right|^2},
\label{c6}
\end{aligned}
\end{equation}
\end{small}with ${\mathcal {A}}= \textsl{g}_{mn}^2{{\eta} _m}(\omega){{\eta _n}}(\omega),$ ${\mathcal {B}}=G_{ab}^2{{\eta} _a}(\omega){{\eta} _b}(\omega),$ ${\mathcal {C}}=\textsl{g}_{mb}^2\\{{\eta} _m}(\omega){{\eta } _b}(\omega)$, and ${\mathcal {D}}= {{\eta} _a}(\omega){{\eta} _b}(\omega){{\eta} _m}(\omega){{\eta} _n}(\omega)$.
The susceptibilities are
\begin{small}
\begin{equation}
\begin{aligned}
{{\eta} _a}(\omega) = &{[ - i(\omega + {\Omega _a}) + {\Gamma_1}/2 + {\Gamma_2}/2]^{ - 1}},\\
{{\eta} _b}(\omega) =&{[ - i(\omega - {\omega_b}) + {\Gamma_3}/2]^{ - 1}},\\
{{\eta} _m}(\omega) =&{[ - i(\omega - {\omega_m}) + {\Gamma_4}/2]^{ - 1}},\\
{{\eta} _n}(\omega) = &{[ - i(\omega - {\omega_n}) + {\Gamma_5}/2 + {\Gamma_6}/2]^{ - 1}}.
\label{c7}
\end{aligned}
\end{equation}
\end{small}When all modes are on resonance, i.e., ${\omega} = - {\Omega _a} = {\omega_b} = {\omega_m} = {\omega_n},$ the frequency conversion efficiency from optics to microwave in Markovian regimes reads
\begin{small}
\begin{equation}
\begin{aligned}
{{\mathcal {U}} _{an}}(\omega) =\frac{{{\Gamma _1}{\Gamma _5}}}{{({\Gamma _1} + {\Gamma _2})({\Gamma _5} + {\Gamma _6})}}\frac{{4{{\vartheta} _{ab}}{{\vartheta} _{mb}}{{\vartheta} _{mn}}}}{{{{[(1 + {{\vartheta} _{ab}})(1 + {{\vartheta} _{mn}}) + {{\vartheta} _{mb}}]}^2}}},
\label{ceq33}
\end{aligned}
\end{equation}
\end{small}where we have defined the photon-phonon cooperativity ${{\vartheta}_{ab}},$ the magnon-phonon cooperativity ${{\vartheta}_{mb}},$ and the magnon-microwave cooperativity ${{\vartheta}_{mn}}$
\begin{small}
\begin{equation}
\begin{aligned}
{{\vartheta} _{ab}} = \frac{{4G_{ab}^2}}{{({\Gamma _1} + {\Gamma _2}){\Gamma _3}}},\,\,
{{\vartheta} _{mb}} = \frac{{4\textsl{g}_{mb}^2}}{{{\Gamma _3}{\Gamma _4}}},\,\,
{{\vartheta} _{mn}} = \frac{{4\textsl{g}_{mn}^2}}{{{\Gamma _4}({\Gamma _5} + {\Gamma _6})}}.
\label{ceq27}
\end{aligned}
\end{equation}
\end{small}Thus, we take into account a configuration where the free variables are the magnon-microwave cooperativity ${{\vartheta} _{mn}}$ and the photon-phonon cooperativity ${{\vartheta} _{ab}},$ which causes the constraint
\begin{small}
\begin{equation}
\begin{aligned}
{{\vartheta} _{ab}} \equiv 1+\frac{{{{\vartheta} _{mb}}}}{{{{\vartheta} _{mn}} + 1}},
\label{ceq35}
\end{aligned}
\end{equation}
\end{small}by fixing ${{\vartheta} _{mb}}$ and setting the partial derivatives of Eq.~(\ref{ceq33}) with respect to ${{\vartheta} _{mn}}$ and ${{\vartheta} _{ab}}$ to zero for the cooperativities. In this case, the maximum efficiency occurs at
\begin{small}
\begin{equation}
\begin{aligned}
{{\vartheta} _{ab}^{max}} = {{\vartheta} _{mn}^{max}}=\sqrt {1 + {{\vartheta} _{mb}}},
\label{ceq351}
\end{aligned}
\end{equation}
\end{small}which leads to the maximum efficiency given by
\begin{small}
\begin{equation}
\begin{aligned}
\max [ {\mathcal {U}}_{an}^{{{\vartheta} _{mb}}}]=\frac{{{\Gamma _1}{\Gamma _5}}}{{({\Gamma _1} + {\Gamma _2})({\Gamma _5} + {\Gamma _6})}}\frac{{(1 + {{\vartheta} _{mb}}){{\vartheta} _{mb}}}}{{{{[1 + {{\vartheta }_{mb}} + \sqrt {1 + {{\vartheta} _{mb}}} ]}^2}}},
\label{cc8}
\end{aligned}
\end{equation}
\end{small}
and the conversion bandwidth satisfying
\begin{small}
\begin{equation}
\begin{aligned}
{{\mathcal {U}} _{an}}({{\omega} _{\max }} \pm \Delta {{\omega} _{1/2}}) =\frac{{{\mathcal {U}} _{an}}( {{\omega} _{\max }})}{2}.
\label{c8}
\end{aligned}
\end{equation}
\end{small}

Finally, we can obtain the Lorentzian resonance spectrum through solving the eigenvalues of matrix $\mathscr{A}$ or matrix $\mathscr{D}$ and bringing them into ${\gamma _{res}}/2{[{(\omega  - {\omega _{res}})^2} + {({\gamma _{res}}/2)^2}]^{ - 1}}$, where the width ${\gamma _{res}}$ of the resonance peak is given by the real part of the eigenvalues, while the imaginary part corresponds to a resonance frequency ${\omega _{res}}$.

\end{document}